\documentclass[useAMS,usenatbib]{mn2e}
\usepackage{graphicx}    
\usepackage{amsmath}
\usepackage{natbib}
\usepackage{color,soul}
\makeatletter
\def\fps@figure{htbp}
\makeatother

\newcommand*{\Msun}{\ensuremath{\mathrm{M_\odot}}}%
\begin{document}
\title[The $K$-band luminosity function]
{Characterising the evolving $K$-band luminosity function using the UltraVISTA, CANDELS and HUDF surveys}
\author[Mortlock et al.]{Alice~Mortlock$^{1}$\thanks{E-mail:
    alicem@roe.ac.uk}, Ross J. McLure$^{1}$\thanks{E-mail: rjm@roe.ac.uk}, Rebecca A. A. Bowler$^{2,1}$, Derek J. McLeod$^{1}$,
\newauthor Esther M\'{a}rmol-Queralt\'{o}$^{1}$, Shaghayegh
Parsa$^{1}$, James S. Dunlop$^{1}$, Victoria A. Bruce$^{1}$ 
\footnotemark[0]\\
$^{1}$Institute for Astronomy, University of Edinburgh, Royal Observatory, Edinburgh
EH9 3HJ\\
$^{2}$Astrophysics, The Denys Wilkinson Building, University of
Oxford, Keble Road, Oxford, OX1 3RH}
\date{}
\pagerange{\pageref{firstpage}--\pageref{lastpage}} \pubyear{2014}
\maketitle
\label{firstpage}

\begin{abstract}
We present the results of a new study of the $K$-band galaxy luminosity
function (KLF) at redshifts $z\leq 3.75$, based on a
nested combination of the UltraVISTA, CANDELS and HUDF surveys.
The large dynamic range in luminosity spanned by this new dataset ($3-4$ dex
over the full redshift range) is sufficient to clearly demonstrate for the
first time that the faint-end slope of the KLF at $z\geq 0.25$ is relatively steep ($-1.3\leq\alpha\leq-1.5$ for a
single Schechter function), in good agreement with recent theoretical and phenomenological models. 
Moreover, based on our new dataset we find that a double Schechter
function provides a significantly improved description of the KLF at $z\leq 2$.
At redshifts $z\geq 0.25$ the evolution of the KLF is remarkably smooth, with little or no evolution evident at faint
($M_{K}\geq-20.5$) or bright magnitudes ($M_{K}\leq-24.5$). 
Instead, the KLF is seen to evolve rapidly at intermediate
magnitudes, with the number density of galaxies at $M_{K}\simeq -23$
dropping by a factor of $\simeq 5$  over the redshift interval
$0.25 \leq z\leq 3.75$. Motivated by this, we explore a simple description of the evolving KLF based on a double Schechter function
with fixed faint-end slopes ($\alpha_{1}=-0.5$, $\alpha_{2}=-1.5$) and a shared characteristic magnitude ($M_{K}^{\star}$).
According to this parameterisation, the normalisation of the component which
dominates the faint-end of the KLF remains
approximately constant, with $\phi^{\star}_{2}$ decreasing by only a factor of $\simeq
2$ between $z \simeq 0$ and $z \simeq3.25$. In contrast, the component which dominates
the bright end of the KLF at low redshifts evolves dramatically, becoming essentially negligible by $z\simeq 3$.
Finally, we note that within this parameterisation, the observed
evolution of $M_{K}^{\star}$ between $z\simeq 0$ and $z \simeq 3.25$ is entirely consistent
with $M_{K}^{\star}$ corresponding to a constant stellar mass of $M_{\star}\simeq 5\times 10^{10}\Msun$ at all redshifts.
\end{abstract}
\begin{keywords}
galaxies: evolution -- galaxies: formation -- galaxies: luminosity function
\end{keywords}

\newpage

\section{Introduction}
\label{sec:int}
As a basic statistical measurement of the galaxy population, the galaxy luminosity function
remains a simple, yet powerful tool for differentiating between competing models of galaxy evolution.
In particular, due to its relative insensitivity to dust reddening and strong correlation with stellar mass,
it has long been recognised that the near-IR luminosity function
provides an insight into the assembly of the underlying stellar mass, without being significantly biased
by recent star formation episodes.

Initial studies of the local $K$-band luminosity function (KLF) were undertaken in the 1990s, thanks to the
rapid developments in near-IR detector technology, but were confined to samples of a few hundred galaxies selected from relatively small areas of sky (e.g. \citealt{Moba93}; \citealt{Glaz95}; \citealt{Gard97}; \citealt{Love00}). This restriction was removed with the arrival of the Two Micron All Sky Survey (2MASS;  \citealt{Jarr00}).
\citet{Koch01} measured the local KLF based on a spectroscopically complete 2MASS sample consisting of 3878 galaxies selected over an area of $\simeq 7000$ deg$^{2}$. 
Contemporaneously, \citet{Cole01} measured the local KLF to fainter magnitudes using a spectroscopic sample of 5683
galaxies within a $\simeq 600$ deg$^{2}$ area of overlap between 2MASS and the 2dF galaxy redshift survey (2dFGRS; \citealt{Coll01}).
The results of the \citet{Koch01} and \citet{Cole01} studies are fully consistent and, after conversion to our adopted cosmology,
indicated that the local KLF could be reasonably well described by a single Schechter function with the following parameters:
$M_{K}^{\star}\simeq-22.5, \phi^{\star}\simeq0.004$~Mpc$^{-3}$, $\alpha\simeq-1.0$.  

Subsequent studies based on combining 2MASS photometry with data from
the Sloan Digital Sky Survey (SDSS; \citealt{Bell03}), the 2dFGRS (\citealt{Eke05})
and the 6dF galaxy survey (\citealt{Jone06}) all measured the local KLF using increasingly large galaxy samples (e.g. 60869 galaxies in the \citealt{Jone06} sample).
Overall the results of these studies are in
reasonable agreement with \citet{Koch01} and \citet{Cole01}, although they typically derive larger number
densities at the bright end ($M_{K}\leq -24$). Moreover, due to the
reduction in statistical errors provided by the larger galaxy samples, it became clear
that a single Schechter function could not simultaneously match the faint and bright
ends of the local KLF in detail.

More recently, \citet{Smit09} studied the local KLF
using a sample of 40111 SDSS galaxies with near-IR photometry
provided by the UKIDSS Large Area Survey (\citealt{Lawr07}). The KLF derived by \citet{Smit09} is in excellent
agreement with that of \citet{Koch01}, but benefits from
significantly reduced statistical errors. Most recently, both \citet{Driv12} and \citet{Kelv14}
measured the local KLF using data from the Galaxy And Mass Assembly survey (GAMA; \citealt{Driv11}; \citealt{Lisk15}). 
Based on a morphological analysis, \citet{Kelv14} found the local KLF to be a composite of Schechter
functions dominated by spheroidal red/passive galaxies and fainter,
bluer, star-forming disk systems respectively (cf. \citealt{Love12}). 
In accord with several previous studies, \citet{Kelv14} found
that a double Schechter function with a
shared value of $M_{K}^{\star}$ offers a significantly improved fit to the
local KLF, being better able to simultaneously fit both the sharp decline at $M_{K}\leq M_{K}^{\star}$
and the upturn seen at faint magnitudes.

In addition to accurately measuring the local KLF, characterising how the KLF evolves
with redshift is clearly important for constraining galaxy-evolution models. \citet{Pozz03} studied the KLF out to $z\simeq 1.5$ using data
covering 52 arcmin$^{2}$ from the K20 survey (\citealt{Cima02}). At their magnitude limit of $K\leq21.9$, \citet{Pozz03} were able to study the KLF at $M_{K} \leq -21$ at $z\simeq 1.0$
and concluded that the KLF primarily displayed luminosity evolution, with $M^{\star}_{K}$ brightening by $\simeq 0.5$ mag between $z=0$ and $z\simeq 1$.
\citet{Pozz03} also highlighted that, even at $z \simeq 1$, the bright end of the KLF is dominated by red/passive objects, and that the
number density of red objects at $z\geq 1$ was significantly under-predicted by contemporary galaxy-evolution models.

This result was confirmed by the wide-area (0.28 deg$^{2}$) study of \citet{Dror03}, based on a sample of
$\simeq 5000$ galaxies with $K\leq 20.6$ and photometric redshifts in
the range $0.4\leq z \leq 1.2$. \citet{Dror03} concluded that
$M^{\star}_{K}$ brightened by $\simeq 0.6$ mag between $z=0$ and $z\simeq 1$, and found an accompanying drop in number density of $\simeq 25\%$.
\citet{Sara06} exploited the ultra-deep $K_{s}$-band imaging in
the HDF-S (FIRES; \citealt{Fran00}) to study the evolution of the KLF down to a magnitude limit of $K_{s}\leq 24.9$; albeit over an area of only 5.5 arcmin$^{2}$.
\citet{Sara06} determined that $M^{\star}_{K}$  brightened  by $\simeq 0.3$ mag between $z=0$ and $z\simeq 1.2$,  in tandem
with a $\simeq 25\%$ drop in number density, in reasonable agreement with both \citet{Pozz03} and \citet{Dror03}.

Based on the original VLT {\sc isaac} $K_{s}$-band imaging of GOODS-S (\citealt{Retz10}), \citet{Capu06} studied the bright
end ($K_{s}\leq 23.4$) of the KLF over the redshift interval $1.0 \leq z \leq 2.5$, using a sample of 2905 galaxies spanning an area of 131 arcmin$^{2}$.
\citet{Capu06} measured a compatible, but somewhat larger, level
of evolution between $z=0$ and $z\simeq 1$, finding that $M^{\star}_{K}$ brightened by $\simeq 0.7$
mag and $\phi^{\star}_{K}$ dropped by a factor of $\simeq 1.5$. 
Over the redshift interval $1<z<2$, \citet{Capu06} highlighted that the number density at the
extreme bright end of the KLF (i.e. $M_{K}<-24$) remains largely
unchanged. Based on this, \citet{Capu06} concluded that the vast
majority ($\simeq85-90\%$) of the most massive galaxies (i.e. $M\geq
2.5\times10^{11}\Msun$; Salpeter IMF) must have already been in
place by $z\simeq 1$, a result which re-enforced the emerging `down-sizing' paradigm.
The results of \citet{Capu06} were confirmed with better statistics by \citet{Cira07}, who used the early release data from the
UKIDSS Ultra-deep survey (UDS; Almaini et al., in preparation) to study the bright end of the KLF over an area of $\simeq 0.6$ deg$^{2}$.

In a later study, \citet{Cira10} addressed the evolution of the KLF based on the UDS DR1, which
provided a sample of $\simeq 50000$ galaxies down to a limit of
$K\leq 23$, and allowed the bright end of the KLF to be traced with
unprecedented accuracy out to $z\simeq 4$. \citet{Cira10} found more clear evidence of down-sizing, finding that
the number density of the brightest galaxies ($M_{K}\simeq -24$) only declines by a factor of $\simeq 2$
from $z \simeq 1$ to $z \simeq 3$, whereas the number density of fainter galaxies  ($M_{K}\simeq-22$) declines by a factor of $\simeq 5$.
Moreover, comparing their KLF measurements to the predictions of galaxy
evolution models demonstrated that all of the contemporary models
appeared to badly over-predict the number density of galaxies
fainter than $M_{K}\simeq-22$. 

Over the last five years, major improvements have occurred, both in
terms of the observational data and the sophistication of the available theoretical predictions.
The primary motivation of this paper is to exploit the latest UV$-$mid-IR
imaging data to provide the most accurate determination yet of the KLF over the redshift interval $0 \leq z \leq 3.75$.
By combining the best available ground-based and space-based imaging
datasets, it is now possible to study the KLF over an unprecedented dynamic range in luminosity ($3-4$ dex over the full redshift range).
This quality of observational data is sufficient to study the evolving form of
the KLF in detail, hence facilitating a meaningful comparison with the
latest generation of galaxy-evolution models.

This paper is set out as follows. In Section 2 we describe
the relevant imaging data before discussing the construction of the 
galaxy catalogues, photometry, and SED fitting in Section 3. In
Section 4 we explain the process of constructing and fitting the
KLF. In Section 5 we present our results on the evolution of the KLF
and compare to previous observational results and the latest
predictions from galaxy-evolution models. In Section 6 we explore a
simple parameterisation for describing the evolution of the KLF before
presenting our conclusions in Section 7. Throughout the paper
magnitudes are quoted in the AB system (\citealt{Oke83}) and we 
assume the following cosmology: $\Omega_{M}=0.3$, $\Omega_{\Lambda}=0.7$ and $H_{0}=70$ km s$^{-1}$ Mpc$^{-1}$.

\section{Data}
\label{sec:data}
For this study we have constructed a final sample of 88484 near-IR selected galaxies within the redshift range
$0.25 \leq z \leq3.75$ from a combination of the UltraVISTA,
CANDELS and HUDF surveys. Together these 
datasets span a factor of $\geq 700$ in terms
of areal coverage and five magnitudes in limiting near-IR depth. The
concatenation of the three individual datasets allows the KLF to be
studied over a dynamic range of $3-4$ dex in luminosity over the full redshift range.
Below we provide a brief description of the datasets and photometry used to construct the final galaxy sample.

\begin{table*}
  \caption{A summary of the depths of the imaging available over the central $\simeq 1$ deg$^{2}$ of UltraVISTA.
    In each case we have listed the median $5\sigma-$depth
  calculated within $2^{\prime\prime}-$diameter circular apertures (or equivalent). 
The $u^{*}g'r'i'z'$ filters refer to the T0007 release of the CFHTLS, the $z'_{2}$ filter
refers to deep Subaru imaging, the $YJHK_{s}$ imaging is from UltraVISTA
and the $3.6\mu$m and $4.5\mu$m imaging is from the {\it Spitzer} SPLASH
survey (see text for full details). Two values are quoted to account for the
  difference in near-IR depth between the deep and wide
  UltraVISTA strips. The depths quoted in the two IRAC bands have been
  corrected to reflect the same fraction of total flux as the optical/near-IR apertures.
It should be noted that the depths of the IRAC imaging display large levels of spatial variation due to the effects of confusion.}
\begin{tabular}{lcccccccccccc}
\hline
\hline
Filter & $u^{*}$ &$g'$ &$r'$& $i'$&  $z'$ &$z_{2}'$ & $Y$& $J$ &$H$ & $K_{s}$ &$3.6\mu$m & 4.5$\mu$m\\
\hline
Deep strip depth & 27.0 & 27.1 & 26.6 & 26.3 & 25.4 & 26.4 & 25.1 & 24.9 & 24.6 & 24.8 & 25.3 & 25.1 \\
Wide strip depth & 27.0  & 27.1  & 26.6  & 26.3  & 25.4  & 26.4  & 24.7 & 24.4 & 24.1 & 23.9 & 25.3  & 25.1 \\          
\hline
\hline
\end{tabular}
\label{tab:depths}
\end{table*}

\subsection{UltraVISTA DR3}
\label{sec:uv_im}
The UltraVISTA survey (\citealt{Mccr13}) images an area of 1.5 deg$^2$ within the Cosmological Evolution Survey (COSMOS) in the
$Y$, $J$, $H$, and $K_{s}$-bands using the Visible and Infrared Camera
(VIRCAM) on the VISTA telescope. The observing pattern employed by UltraVISTA results in half of the total area being
covered by ultra-deep strips (referred to here as UltraVISTA deep), with the other half
covered by shallower inter-strip regions (referred to here as UltraVISTA wide).

In this study we employ the latest DR3 release of the UltraVISTA dataset, and specifically
utilise the 1 deg$^2$ overlap region with the $u^{*}$, $g'$, $r'$, $i'$,
and $z'$-band imaging provided by the T0007 release of the Canada-France-Hawaii Telescope Legacy Survey (CFHTLS;  \citealt{Hude12}).
Within this 1 deg$^2$ region, $\simeq60\%$ is covered by the UltraVISTA wide strips with a 5$\sigma-$limit of $K_{s}\!=$23.9 (2$^{\prime\prime}-$diameter
aperture). The remaining $\simeq40\%$ of the region is covered by UltraVISTA deep strips which have a 5$\sigma-$limit of $K_{s}\!=$24.8.
In addition, we also exploit new deep $z'$-band imaging covering the UltraVISTA region taken with Suprime-Cam on Subaru (\citealt{Furu08}).
Finally, we utilise deep imaging of the COSMOS field at 3.6$\micron$
and 4.5$\micron$ taken with \textit{Spitzer}/IRAC. This data comes from a combination of the
\textit{Spitzer} Extended Deep Survey (SEDS; \citealt{Ashb13}) and the
\textit{Spitzer} Large Area Survey with Hyper-SuprimeCam (SPLASH; PI:
Capak, \citealt{Stei14}).

All images from the $u^{*}$ to $K_{s}$-band were resampled onto the
same pixel scale (0.186 arcsec/pixel) and shifted to the same zero
point (see \citealt{Bowl14} for a detailed discussion). In addition, all
images from the $u^{*}$ to $K_{s}$-bands were PSF homogenised to match
the $Y$-band image which has the poorest seeing of the optical/near-IR
data (FWHM=$0.8^{\prime\prime}$). A summary of the imaging data  available over the central $\simeq 1$ deg$^{2}$ of UltraVISTA is provided in Table \ref{tab:depths}.

\subsection{CANDELS}
\label{sec:can_im}
The Cosmic Assembly Near-infrared Deep Legacy Extragalactic Survey (CANDELS) provides optical and
near-IR {\it Hubble Space Telescope} ({\it HST}) imaging over an area of
$\simeq 0.25$ deg$^2$, divided between five different survey fields
(\citealt{Grog11}; \citealt{Koek11}). In this work we utilise the
CANDELS data available in the Ultra Deep Survey (UDS) and GOODS-S fields.

\subsubsection{CANDELS/UDS}
\label{sec:cuds_im}
The CANDELS data in the UDS field covers an area of $\simeq 0.06$ deg$^2$ and
consists of F606W ($V_{606}$) and F814W ($I_{814}$) optical imaging
taken with ACS and F125W ($J_{125}$) and F160W ($H_{160}$) near-IR imaging taken with WFC3/IR.
The CANDELS/UDS field is a sub-set of the full UKIDSS UDS field, which covers an area of $\simeq 0.8$ deg$^2$ with deep
ground-based imaging in the $J$, $H$ and $K$-bands. 
In addition, the CANDELS/UDS area is covered by deep $U$-band data from
the {\it CFHT} and deep optical imaging in the  $B,V,R,i^{\prime}$ and $ z^{\prime}$-bands from the Subaru/XMM-Newton Deep Survey (SXDS; \citealt{Furu08}).
Moreover, the CANDELS/UDS region is covered by deep $Y$ and
$K_{s}$-band imaging taken with HAWK-I on the {\it VLT} as part of the HUGS survey (\citealt{Font14}).
Finally, CANDELS/UDS is also covered by deep 3.6$\mu$m and 4.5$\mu$m {\it Spitzer} IRAC imaging from SEDS and SCANDELS (\citealt{Ashb15}).

\subsubsection{CANDELS/GOODS-S}
\label{sec:cgds_im}
The CANDELS data in the GOODS-S field covers an area of $\simeq 0.05$
deg$^2$ and consists of optical ACS imaging in the F606W ($V_{606}$)
and F814W ($I_{814W}$) filters and WFC3/IR near-IR imaging in the F105W
($Y_{105}$), F125W ($J_{125}$) and F160W ($H_{160}$) filters \citep{Koek11}. 

The CANDELS WFC3/IR imaging in GOODS-S is split into two distinct
regions (deep and wide) which received 5 orbits and 2 orbits of
near-IR observations respectively (with the ACS imaging obtained in parallel).
In addition, the northern third of the GOODS-S field is covered
by F098M ($Y_{098}$), F125W ($J_{125}$) and F160W ($H_{160}$) imaging
taken as part of the WFC3/IR Early Release Science (ERS) programme \citep{Wind11}.
The CANDELS GOODS-S field is also covered by deep ACS
imaging in the F435W ($B_{435}$), F606W ($V_{606}$), F775W ($I_{775}$)
\& F850LP ($z_{850}$) filters taken as part of the original GOODS programme \citep{Giav04}. The HUDF region, which
features the deepest ACS and HUDF imaging available (\citealt{Bouw09}; \citealt{Elli13}; \citealt{Koek13}), is situated within the GOODS-S deep region.

In addition to the {\it HST} imaging, the GOODS-S field is covered by a
large amount of ancillary ground-based imaging data. Of particular importance to
this study are the ultra-deep $U$ and $K_{s}$-band imaging taken with VIMOS
\citep{Noni09} and HAWK-I (\citealt{Font14}) respectively.
Finally, the GOODS-S field is covered by ultra-deep {\it Spitzer} IRAC
imaging taken as part of the original GOODS programme (PI: Dickinson), SEDS and SCANDELS.

\subsection{UltraVISTA photometry}
\label{sec:uv_phot}
To measure the optical$-$near-IR photometry in UltraVISTA,
\textsc{sextractor} (version 2.8.6; \citealt{Bert96}) was run in dual-image 
mode on the PSF-matched images, using the UltraVISTA
$K_{s}$-band mosaic as the detection image. The basic photometry was measured within
$2^{\prime\prime}$-diameter circular apertures, with photometric errors computed on
an object-by-object basis using measurements of the local image depths.
For a given object, the local depth in a given image was measured from the
aperture-to-aperture variance of the closest 200 blank-sky
apertures. In this procedure the variance was calculated using the
robust Median Absolute Deviation (MAD) estimator and the blank-sky
apertures were drawn from a grid defined for each image using the
appropriate  \textsc{sextractor} segmentation map.
For the purposes of SED fitting, the photometric errors for bright objects
were forced to be $\geq5\%$, in order to reflect systematic uncertainties
in the zero-point calibration and aperture corrections.

As a result of the comparably poor spatial resolution of the {\it
  Spitzer} IRAC imaging (FWHM $\simeq 1.7^{\prime\prime}$), blending of
nearby objects means that it is not possible to extract reliable
photometry using \textsc{sextractor} in dual-image mode. 
Consequently, in order to obtain accurate photometry in the 3.6$\mu$m
and 4.5$\mu$m bands the deconfusion code \textsc{tphot} was employed \citep{Merl15}. 
\textsc{tphot} is an updated and improved version of
the \textsc{tfit} code \citep{Laid07}, and uses the positions and
morphologies of objects measured in a high-resolution image (in this case the $K_{s}$-band) as prior
information to simultaneously solve for the corresponding fluxes in a
low-resolution image (in this case the $3.6\mu$m and $4.5\mu$m IRAC imaging).
In order to reflect the systematic problems associated with deconfused
photometry, the covariance matrix flux errors delivered by \textsc{tphot} were set to a
minimum level of 10\% for bright objects, in order to avoid the deconfused
fluxes for such objects being associated with unrealistically high signal-to-noise levels.

\subsection{CANDELS photometry}
\label{sec:can_phot}
For the purposes of this study we adopted the photometry catalogues of
the CANDELS UDS and GOODS-S fields publicly released by the CANDELS team. 
Below we briefly describe the key elements of how these catalogues were produced, but a full description of the production of
the public UDS and GOODS-S photometry catalogues is provided by \citet{Gala13}
and \citet{Guo13} respectively.

Object detection was performed with \textsc{sextractor} using
the $H_{160}$ mosaics as the detection images. \textsc{sextractor} was run
in a `{\it hot}' and `{\it cold}' configuration to optimise detection of compact
and extended sources respectively. All other {\it HST} imaging was
PSF-homogenised to the $H_{160}$ imaging and photometry was extracted
by running \textsc{sextractor} in dual image mode. The {\it HST}
fluxes were initially measured as \textsc{flux\_iso} before being converted
to total fluxes using aperture corrections based on the ratio of
\textsc{flux\_iso} to either \textsc{flux\_best} (\citealt{Gala13}) or \textsc{flux\_auto} (\citealt{Guo13}) in the $H_{160}$-band. Photometry was extracted from the lower spatial resolution
ground-based and {\it Spitzer} imaging with the \textsc{tfit}
deconfusion code (\citealt{Laid07}) using the $H_{160}$-band imaging
as prior information.

\begin{figure*}
\includegraphics[trim = {25mm 130mm 20mm 60mm}, clip,width=8.5cm]{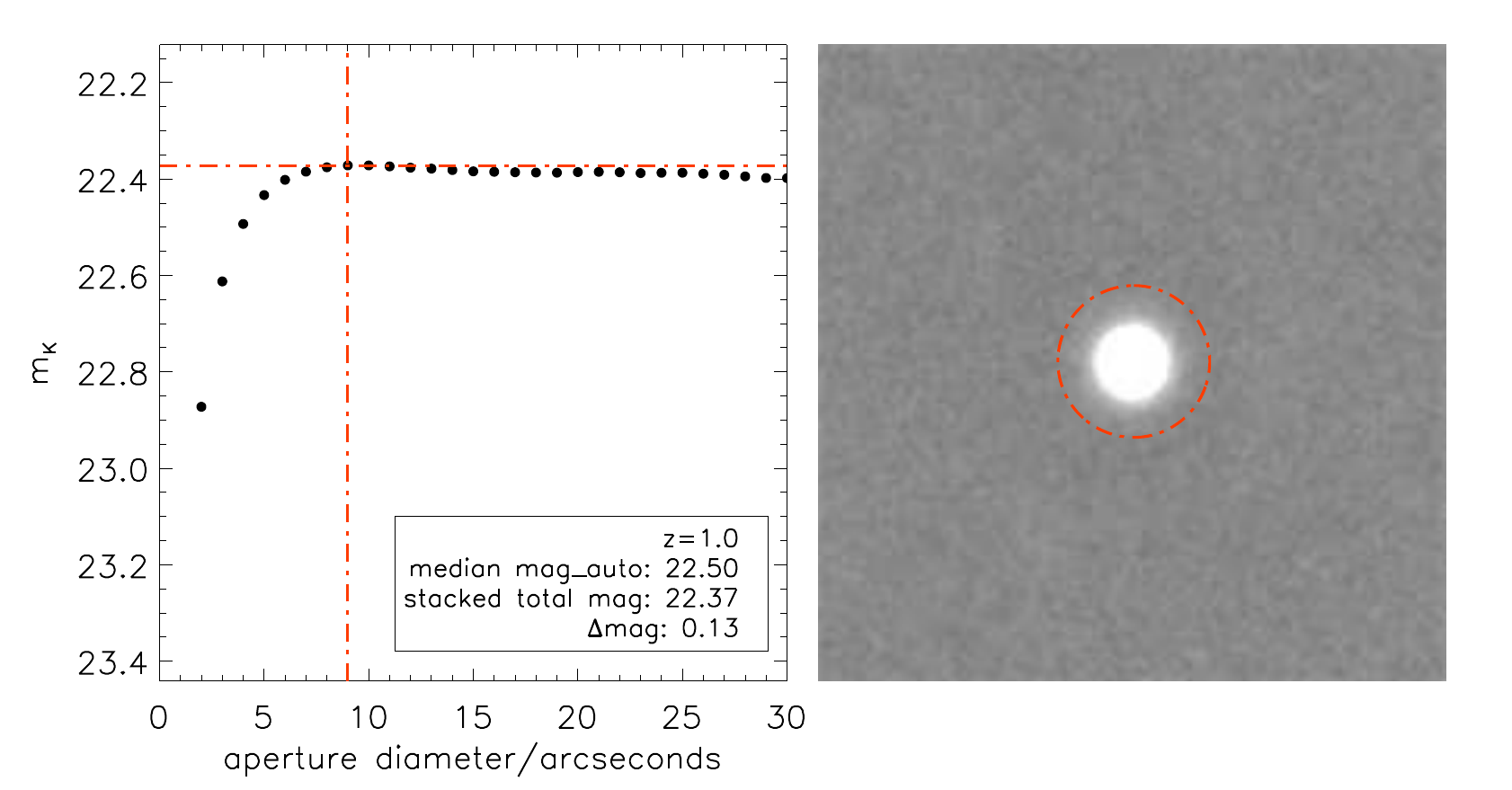}
\includegraphics[trim = {25mm 130mm 20mm 60mm}, clip,width=8.5cm]{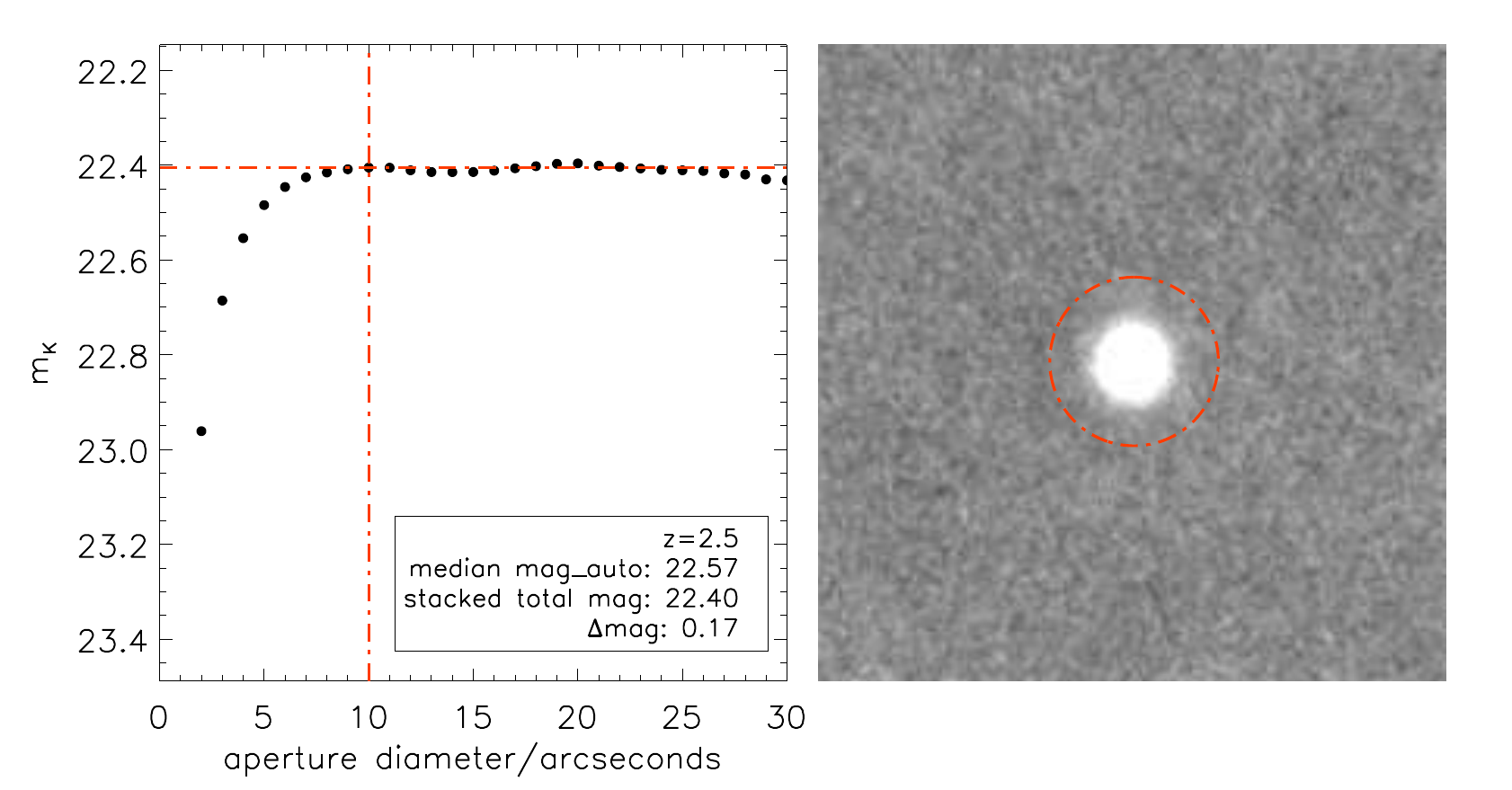}
\caption{Examples of the stacking analysis used to determine the best
  correction between $2^{\prime\prime}-$diameter aperture fluxes and total
  fluxes at $z=1$ (left) and $z=2.5$ (right).
  In both cases the
  right-hand sub-panel shows the stacked image of galaxies within the
  \textsc{mag\_auto} range $22<m_{\rm auto}<23$.
  The left-hand sub-panels show the
  corresponding curve-of-growth, where the
  horizontal red dashed line shows the best estimate of the total magnitude. The vertical red dashed line shows the
  aperture diameter that encloses the total flux, which can be seen as the red circular aperture in the right-hand sub-panels.
  The inset legends show the value of $\Delta$mag between the curve-of-growth estimate of total magnitude and the median \textsc{mag\_auto} of those galaxies entering the stack.}
\label{cog_plot}
\end{figure*}

\subsection{Correction to total magnitude}
\label{sec:total_mag}
The UltraVISTA photometry was initially measured in $2^{\prime\prime}-$diameter circular
apertures before being corrected to total magnitudes based on the
ratio of \textsc{flux\_auto} to \textsc{flux\_aper} in the $K_{s}$-band. 
This is a fairly standard approach to correcting to total magnitudes, 
and relies on the assumption that the Kron-like magnitudes (Kron 1980) measured by
\textsc{sextractor} (i.e. \textsc{mag\_auto}) capture $\simeq 90\%$ 
of the integrated galaxy light (depending on the form of the radial surface-brightness profile).
However, the different techniques commonly adopted to correct to total
magnitudes have been the subject of extensive discussion in the recent
literature, with variations in the methods adopted for bright objects being
blamed for inconsistencies between different determinations of the bright end of
the galaxy luminosity function (e.g. \citealt{Bern13}; \citealt{DSou15}; \citealt{Love15}). 

In order to address this issue, we produced stacked
$K_{s}$-band images of our final galaxy sample as a function of
redshift and apparent $K_{s}$-band magnitude. To ensure that flux from
nearby bright companions did not mimic extended wings to the
stacked radial surface-brightness profiles, the stacking was confined
to the 65$\%$ of the final $K_{s}$-band galaxies with no companions
within a radius of $4^{\prime\prime}$ with fluxes greater than 50\% of
the flux of the primary object. Postage-stamp $K_{s}$-band images of each galaxy were generated,
sky-subtracted and cleaned of nearby companion objects. 
Each postage-stamp image was initially sky-subtracted using the median
of all sky pixels at radii $\geq 5^{\prime\prime}$ (as indicated by the
\textsc{sextractor} segmentation map). Additionally, a further level
of sky-subtraction was applied by fitting a two-dimensional surface
(first order polynomial) to all sky pixels, excluding all pixels within a radius
of 7.5$^{\prime\prime}$ from the central object.

After this cleaning process, median stacks of the objects in each
redshift and apparent magnitude bin were produced. When constructing
the median stacks, all pixels identified by \textsc{sextractor} as
belonging to companion objects were excluded. A non-parametric
measurement of the total flux in each stacked image was then derived using a curve-of-growth analysis. Example curve-of-growth plots can be seen in Fig. \ref{cog_plot}.

The results of the curve-of-growth analysis demonstrated that the
total flux recovered from the stacked images was systematically larger
than the median value of \textsc{flux\_auto} for the objects included in the
stack. Interestingly, the off-set between the total stacked flux and
the median \textsc{flux\_auto} of the stacked galaxies varied very little with
either redshift or apparent $K_{s}$-band magnitude. The median offset was 13.5\%, with the off-set always lying within the range $9-18\%$. Consequently, throughout the rest of this analysis we
adopt 1.135$\times$\textsc{flux\_auto} as our best estimate of the
total flux for the UltraVISTA galaxies. Based on an identical stacking analysis of the
$H_{160}$ imaging in the two CANDELS fields, no systematic
off-set was found between the curve-of-growth fluxes and the total fluxes determined by \citet{Guo13} and \citet{Gala13}. As a consequence, no correction was applied to the CANDELS photometry.

As part of this analysis, a stack of several hundred isolated stars was
used to accurately measure the correction between $2^{\prime\prime}-$diameter magnitudes and total magnitudes for point sources in the PSF homogenised UltraVISTA data. 
The resulting correction of $-0.43$ magnitudes was then used as a {\it
  minimum} correction for all galaxies, i.e. no object was assigned a correction smaller than this.
   The minimum correction was applied in order to ensure that no
galaxies were corrected by an amount less than expected for an unresolved point source.

\section{Sample selection}
In this section we describe the process of constructing the final sample of $0.25 ~\leq~z~\leq~3.75$ galaxies, based on spectral energy distribution (SED) fitting of the photometry catalogues described above. 
The upper limit to the redshift interval was specifically chosen to provide three photometric data points sampling the galaxy SED long-ward of the $4000$\AA\, break
(i.e. $K_{s}$, $3.6\mu$m \& $4.5\mu$m), with the {\it Spitzer} IRAC photometry always
providing a measurement of the rest-frame SED at $\lambda_{\rm rest} \geq 1\mu$m. This situation naturally limits the uncertainties involved in
calculating the absolute $K$-band magnitude for each galaxy. 

In addition to the SED fitting process, we also describe the simulations performed in order
to calculate completeness corrections and how the galaxy sample was
cleaned of galactic stars, artefacts and potential active galactic nuclei (AGN).

\subsection{Star-galaxy separation}
Before calculating photometric redshifts, the galaxy sample was first
cleaned of galactic stars. Within the UltraVISTA dataset, this was initially performed
by removing objects lying on the stellar locus in the ($Y-J$) vs
($H-K$) colour-colour diagram. As a secondary check, objects consistent with the stellar locus on the
($g'-i'$) vs~($J-K$) colour-colour diagram were also removed (see \citealt{McCr12} and \citealt{Jarv13} for examples of star-galaxy separation in data sets of this size). 
Due to their small area, the stellar loci in the CANDELS datasets are poorly defined. However, due to the high-spatial resolution of the
{\it HST} imaging, star-galaxy separation is relatively straightforward. 
Consequently, we removed all point-like objects with \textsc{sextractor}
parameter \textsc{class\_star}$>$0.98, a threshold demonstrated to efficiently isolate objects with stellar colours by \citet{Gala13}.

\begin{figure*}
\includegraphics[trim = {35mm 180mm 30mm 40mm}, clip,width=17.0cm]{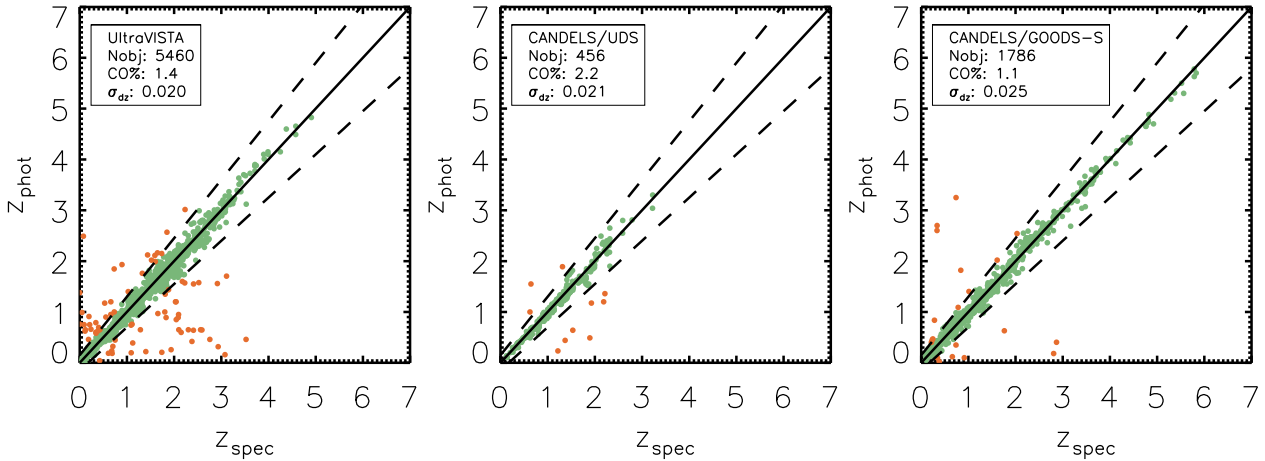}
\caption{A comparison of spectroscopic and photometric redshifts within the final galaxy sample. The left-hand panel shows $z_{phot}$ vs $z_{spec}$ for UltraVISTA deep and wide combined, whereas the middle and right-hand panels show the equivalent information for the CANDELS/UDS and CANDELS/GOODS-S fields respectively. The black solid lines show the one-to-one relation and the black dashed lines show the cut used for identifying catastrophic outliers (i.e. $\left| dz \right|>0.15$). Catastrophic outliers are plotted in orange, while those objects with acceptable photometric redshifts are plotted in green. The inset panels list the basic statistics of the comparison in each field: number of spectroscopic objects ($N_{obj}$), percentage of catastrophic outliers (CO\%) and $\sigma_{dz}$.}
\label{specz_vs_photoz}
\end{figure*}

\subsection{SED fitting}
\label{sec:sed_fitting}
Following the removal of galactic stars, photometric redshifts were
computed using template-based SED fitting to the PSF-matched photometry
catalogues described in Section 2. The final photometric redshifts
adopted for the analysis were the median of five different estimates
produced using different codes and different template sets. The different photometric redshift code+template configurations were as follows:
\begin{enumerate}
\item {Two sets of photometric redshifts were generated using the
    publicly available SED fitting code \textsc{lephare} (\citealt{Arno99}; \citealt{Ilbe06}), both assuming solar metallicity, the \citet{Calz00} dust
    attenuation law with E(B-V) values in the range $0.0-0.5$ and
    including emission lines. The first photometric redshift run employed the \textsc{zcosmos} template 
set \citep{Ilbe09} while the second employed the \textsc{p\'{e}gase}
template set.}
\item{A further two sets of photometric redshifts were generated using
    the publicly available \textsc{EAZY} code (\citealt{Bram08})
    using the PCA (\citealt{Blan07}) and \textsc{p\'{e}gase}
    template sets (\citealt{Fioc99}).}
\item{The final set of photometric redshift results was generated
    using a private SED fitting code (\citealt{McLu11}; \citealt{McLe16}) employing \citet{Bruz03} (BC03) templates with
    metallicities in the range $0.2Z_{\odot}-Z_{\odot}$ and the
    addition of strong emission lines. This photometric redshift set-up employed the \citet{Calz00} dust attenuation law,
allowing $A_{V}$ to vary within the range $0.0\leq A_{V} \leq2.5$. }
\end{enumerate}

A comparison between our final median photometric redshifts and
publicly available spectroscopic redshifts is shown in
Fig. \ref{specz_vs_photoz}. It can be seen that our photometric redshift
results have a typical value of $\sigma_{dz}\simeq 0.02$ (calculated
using the MAD estimator), where $dz=z_{spec}-z_{phot}/(1+z_{spec})$. Using the standard definition
of catastrophic outliers as those objects with $\left| dz \right| >0.15$, the typical
catastrophic outlier rate is $\simeq 1-2\%$. These results indicate that
our photometric redshifts are robust and do not vary in quality between the space-based and ground-based photometry.

In order to calculate the final values of absolute $K$-band magnitude
($M_{K}$), the SED of each galaxy was fit for a final time (at fixed redshift) using a sub-set of BC03 templates defined by \citet{Wuyt11b}. 
This set of templates consists of exponentially decaying star-formation histories with values of $\tau$ in the range $0.3 \leq \tau \leq10$ Gyr and ages in the range 50 Myr to the age of the Universe.
Dust extinction was applied using the \citet{Calz00} attenuation law
with E(B$-$V) allowed to vary within the range $0.0-0.5$ and metallicity
was fixed at solar.

The advantage of this template sub-set is that it produces
star-formation rate (SFR) estimates which are in good agreement with estimates of
total SFRs calculated from the addition of raw UV star
formation and dust obscured star formation measured at sub-mm
wavelengths \citep{Wuyt11b}. However, it should be noted that, given a fixed redshift and
multi-wavelength data covering the observed wavelength range
$0.38\mu$m $ \leq \lambda \leq 4.5\mu$m, the derived values of absolute $K$-band
magnitude are not very dependent on the assumed SED template set. The
distribution of absolute $K$-band magnitude versus redshift for
the final sample is shown in Fig. \ref{redshift_vs_mk}, where the
values of $M_{K}$ have been corrected to total according to the
prescription described in Section 2.5.

\subsection{Completeness}
\label{sec:comp}
When measuring the KLF it is vital to accurately calculate the
completeness limits of the data, particularly when trying to measure the faint-end slope. 
In order to compute the completeness, a synthetic galaxy population was
created in each field, covering the redshift range $0.25 \leq z \leq3.75$ and the appropriate range in $M_{K}$ for that dataset.
The number densities as a function of magnitude were based on the
\citet{Cira10} parameterisation of the evolving KLF although, given
our final choice of conservative cuts (see below), the exact input KLF
parameters do not have a significant impact on the final completeness corrections.

Each member of the synthetic galaxy population was randomly allocated
an SED template taken from the catalogue of SED fits to members of the
real galaxy sample, matched within $\Delta z = \pm 0.25$ and $\Delta M_{K}=\pm 0.25$ magnitudes.
Based on the adopted SED template, the synthetic galaxy was injected
as a point source into the relevant UltraVISTA or CANDELS imaging data
with the appropriate $K$-band or $H_{160}$-band apparent magnitude.
The completeness as a function of apparent magnitude and redshift was
then calculated by analysing the images containing the synthetic
sources with an identical \textsc{sextractor} configuration to that
employed when selecting the original samples. See Fig. \ref{redshift_vs_mk} for the completeness of each survey 
as a function of redshift.

\subsection{Final cleaning}
\label{sec:samp_selec}
Before proceeding to measure the KLF, the galaxy sample was cleaned of
objects with erroneous photometry and potential AGN. The first stage in
this process was to remove the 5\% of objects with the poorest quality
SED fits as indicated by their $\chi^{2}$ values. Given the
anti-correlation between redshift and $\chi^{2}$, this cleaning was
done separately within the six redshift bins adopted for the rest of the analysis.
The vast majority of objects excluded on the basis of their high
$\chi^{2}$ were either artefacts or objects whose photometry was contaminated/corrupted in one or more filters.

The second stage in the final cleaning process was to exclude
potential AGN. Within the UltraVISTA dataset, all sources were removed which were detected in either the Chandra Cosmos
survey (C-COSMOS; \citealt{Elvi09}) or the VLA-COSMOS Large Project
(\citealt{Schi07}). Finally, we also used the S-COSMOS 24$\mu$m catalogue \citep{Sand07} to
remove potential obscured AGN. This was achieved by converting the
24$\mu$m flux into a measurement of specific SFR (SSFR) using the \citet{Riek09} IR templates and stellar masses computed using the same SED templates as in Section \ref{sec:sed_fitting}.
The 24$\mu$m SSFR distribution shows a clear bi-modality with the high SSFR peak being defined as log$_{10}(SSFR /\rm Gyr^{-1})>0.75$.
The peak is dominated by $z>2$ objects where the 24$\mu$m is sampling hot dust which is likely AGN heated. We therefore remove all objects within UltraVISTA in the high peak of this distribution in the redshift range $0.25\leq z \leq 3.75$. The combination of both AGN cleaning methods resulted in removal of 7\% of the sample across the full redshift range. Within the CANDELS/UDS dataset we excluded potential AGN on
the basis of the X-ray/radio detections provided in the publicly
available photometry catalogue \citep{Gala13}. For the CANDELS/GOODS-S dataset, potential AGN were excluded
by matching to the X-ray, radio and IR-selected AGN candidates
compiled by \citet{Koce12}. 

\begin{figure}
\centering
\includegraphics[trim = 22mm 122mm 0mm 22mm, clip,scale=0.6]{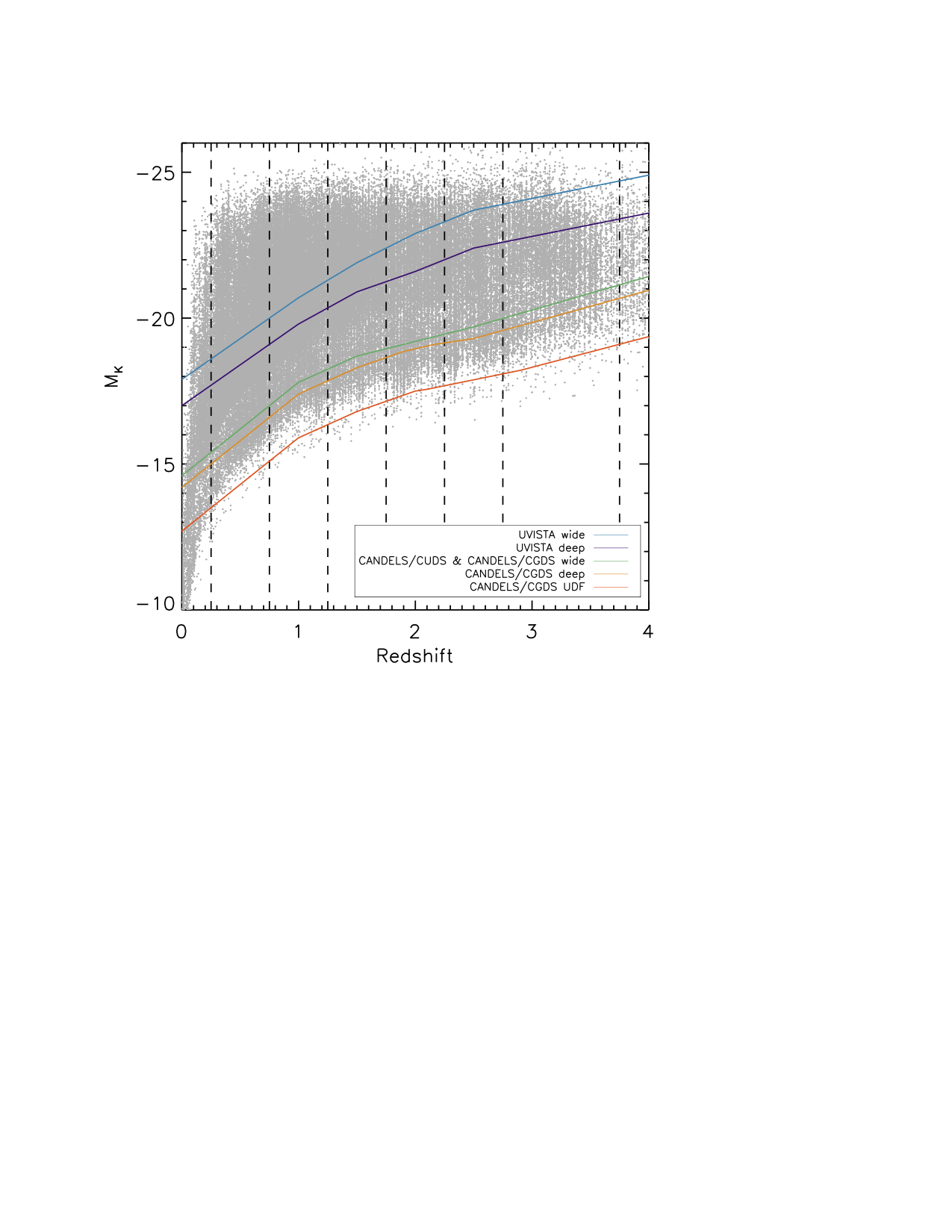}
\caption{Redshift versus absolute $K$-band magnitude for the final galaxy sample (grey data points). The black vertical dashed lines show the limits of the redshift bins employed in the rest of the analysis,
  and the coloured solid lines are the 95\% completeness limits of the five different surveys used in this work.}
\label{redshift_vs_mk}
\end{figure}

After completing the final cleaning process it was possible to
construct the final galaxy sample to be used in the computation of the KLF. 
Fortunately, the different datasets used to construct the final galaxy
sample cover ranges in $K$-band luminosity which overlap significantly. As a
result, it was possible to adopt a conservative approach and only
include a galaxy in the final sample if it survived the full cleaning
process and was brighter than the 95\% completeness limit for its redshift
within the survey from which it was originally selected. The
distribution of the final galaxy sample on the $M_{K}-z$ plane is
shown in Fig. \ref{redshift_vs_mk}, along with the 95\% completeness
limits for the various different surveys.

\section{Luminosity function fitting}
Armed with the final galaxy sample, the KLF was derived using
the classical $\frac{1}{V_{max}}$ (Schmidt 1968) maximum likelihood estimator, defined as follows:
\begin{equation}
\phi(M_{K}) \Delta M = \sum\limits_{i=1}^{N_{gal}} \frac{1}{C(M_{K},z)V_{max,i}} 
\label{eq:vmax}
\end{equation}
\noindent
where $\phi(M_{K})$ is the number density in a given absolute $K$-band
magnitude bin, $V_{max,i}$ is the maximum volume a given object can be
associated with and still be included within the sample, $C(M_{K},z)$ is
the completeness as a function of absolute $K$-band magnitude and 
redshift (as computed in Section \ref{sec:comp}) and $\Delta M$ is the bin size in magnitudes.

The full uncertainties associated with the number densities are a combination of Poisson
error ($\sigma_{poi}$), error due to template fitting ($\sigma_{temp}$), and cosmic variance
($\sigma_{cv}$). The Poisson contribution is computed as:
\begin{equation}
\sigma_{poi} = \sum\limits_{i=1}^{N_{gal}} \sqrt{\frac{1}{C(M_{K},z)V_{max,i}}}
\label{eq:poi_err}
\end{equation}
\noindent
The template fitting error ($\sigma_{temp}$) is computed from Monte Carlo simulations which account for the uncertainties introduced from
the errors in object photometry and potential mismatches between the
real galaxy SEDs and the adopted model templates. In this process, 100
realisations of the KLF across all redshift bins were computed, with
each individual galaxy randomly allocated a redshift, drawn from its
P(z) distribution, and allocated updated values of $M_{K}$ and completeness.
The value of $\sigma_{temp}$ in a given redshift and $M_{K}$ bin was
then calculated from the distribution of number densities returned by the Monte Carlo realisations.
As described previously, the final adopted 
redshift for each galaxy was the median
of five separate photometric redshift runs, each of which delivered a $1\sigma$ upper and lower confidence region (i.e. $z_{min}<z_{phot}<z_{max}$).
Consequently, the P(z) for each object was modelled as a two-sided Gaussian function, centred on the median redshift ($z_{med}$) with
the upper and lower sigma values set to  $\sigma_{high}=z_{max}-z_{med}$ and $\sigma_{low}=z_{med}-z_{min}$ respectively. In order to be
conservative, the values of $z_{max}$ and $z_{min}$ adopted to
construct the P(z) distributions were the extreme values returned by
the five photometric redshift runs.

Finally, we compute the value of $\sigma_{cv}$ at a given redshift and within a
given $M_{K}$ bin. Given that our final dataset is effectively the combination of five different surveys, we are able to use the variation in galaxy number density within a given $M_{K}$ and redshift bin as an empirical measurement of the cosmic variance uncertainties. The final uncertainty on
the number density calculated at a given redshift and within a given
$M_{K}$ bin was then taken as:
\begin{equation}
\sigma_{\phi(M_{K})}= \sqrt{(\sigma_{poi}^2 + \sigma_{temp}^2 + \sigma_{cv}^2)}
\label{eq:total_err}
\end{equation}

In our faintest magnitude bins, where we only have data from the UDF, we cannot estimate the cosmic variance
uncertainty using our standard method, and make the assumption that Poisson and template uncertainties are dominant.
To test this assumption we estimated the cosmic variance in these bins according to the prescription of \citet{Most11}.
This calculation suggests that the total error in the faintest bins would likely be increased by only $\sim 20$\%, confirming that the Poisson and
template uncertainties are dominant. Fitting the KLF data with and without this additional contribution results in best-fitting parameters
and uncertainties which are virtually identical to those presented in Table \ref{tab:LF_params}.

\subsection{Schechter-function fits}
\label{sec:sch_func}
Throughout the analysis we employ $\chi^{2}$ fitting to the binned KLF
data using either single or double Schechter functions. The single Schechter function has the following form:
\begin{equation}
\phi(M) = 0.92\phi^{*}\cdot(10^{-0.4(M-M^{*})})  ^{(1+\alpha)} e^{[-10^{-0.4(M-M^{*})}]}
\label{eq:single_sch_LF}
\end{equation}
\noindent where $\phi^{*}$ is the normalisation, M$^{*}$ is the characteristic magnitude and $\alpha$ is the faint-end slope.
The double Schechter function is parameterised as follows:
\begin{multline}
\phi(M) = 0.92\cdot10^{-0.4(M-M^{*})} e^{[-10^{-0.4(M-M^{*})}]}\\
\cdot [\phi_{1}^{*}\cdot10^{-0.4(M-M^{*})\alpha_{1}}+\phi_{2}^{*}\cdot10^{-0.4(M-M^{*})\alpha_{2}}]   
\label{eq:double_sch_LF}
\end{multline}
\noindent where M$^{*}$ is the shared characteristic magnitude and
($\phi_{1},\alpha_{1}$) and ($\phi_{2},\alpha_{2}$) are the
normalisations and faint-end slopes of the two Schechter-function
components. 

\subsection{The local $K$-band luminosity function}
\begin{figure}
\centering
\includegraphics[trim = 22mm 122mm 65mm 35mm, clip, width=8.5cm]{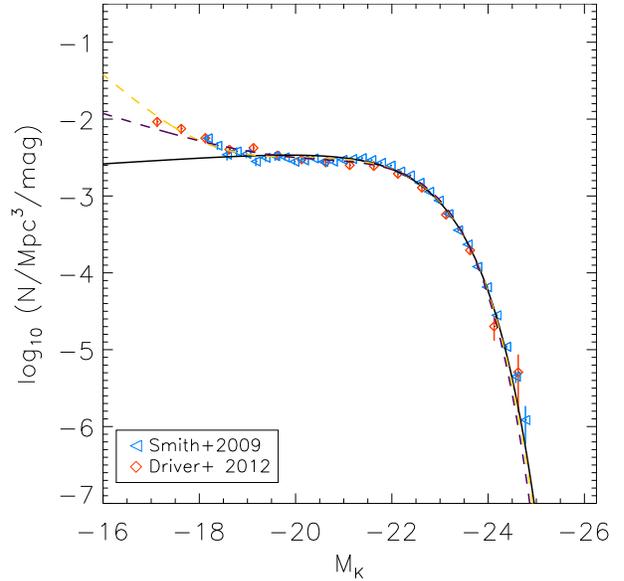}
\caption{The local $K$-band galaxy luminosity function as measured by \citet{Smit09} and \citet{Driv12}.
The solid black and dashed yellow lines are our best-fitting single and double Schechter-function fits to the combined dataset respectively.
The dashed purple line shows our best-fitting double Schechter function with fixed faint-end slopes of $\alpha_{1}=-0.5$ and $\alpha_{2}=-1.5$ (see Section 6 for a discussion).}
\label{KLF_local}
\end{figure}

Before proceeding to explore the evolution of the KLF, it is clearly desirable to have a robust measurement of the local KLF to serve as a baseline. In Fig. \ref{KLF_local} we show the local KLF based on data from the UKIDSS LAS (\citealt{Smit09}) and GAMA surveys (\citealt{Driv12}). The KLF data from \citet{Smit09} was converted to total magnitudes assuming $K_{AB}=K_{vega}+1.9$ and $K_{\rm tot}-K_{\rm Petrosian}=-0.2$, whereas the KLF data from \citet{Driv12} was converted to total magnitudes assuming $K_{\rm tot}-K_{\rm Kron}=-0.1$.

After making the necessary corrections, it can be seen from
Fig. \ref{KLF_local} that the \citet{Smit09} and \citet{Driv12} datasets are completely compatible and, as a result, a
combined fit to both datasets was performed in order to derive our fiducial local KLF
parameters. In Fig. \ref{KLF_local} the best-fitting single and double
Schechter-function fits are shown as the solid black and
dashed yellow lines respectively. The dashed purple line shows the best-fitting double Schechter function with faint-end slopes fixed at $\alpha_{1}=-0.5$ and $\alpha_{2}=-1.5$ (see Section 6 for a discussion). The best-fitting parameters corresponding to the three fits shown in Fig. \ref{KLF_local} are listed in Table \ref{tab:local}

As might be expected, our single Schechter-function fit to the combined local KLF data is intermediate to those derived by \citet{Smit09} and \citet{Driv12}, and
is in good agreement with previous fits reported by \citet{Koch01} and \citet{Cole01}. It can be seen from Fig. \ref{KLF_local} that the local KLF is reasonably
well matched by a single Schechter function at magnitudes brighter than $M_{K}\simeq -20$, with convincing evidence for an up-turn in the number density of galaxies only
apparent within the faintest few magnitude bins. Formally the best-fitting double Schechter function has a very steep faint-end slope ($\alpha_{2}=-2.35\pm0.30$),
although the lack of dynamic range in luminosity means that the slope and normalisation of this component are not very well constrained. The uncertainty in the slope of the
second Schechter function component can clearly be seen in Fig. \ref{KLF_local},
where the best-fitting double Schechter function (yellow dashed line) is virtually indistinguishable from the double Schechter-function fit with the faint-end slope fixed at $\alpha_{2}=-1.5$ (purple dashed line).

\begin{table*}
\caption{The best-fitting single and double Schechter-function
  parameters for the local KLF dataset shown in
  Fig. \ref{KLF_local}. Columns $2-6$ list the best-fitting parameters
  and their corresponding uncertainties. Columns 7 \& 8 list the
  corresponding values of $\chi^{2}$ and reduced $\chi^{2}_{\nu}$
  respectively. The final row shows the results of fitting the local
  KLF dataset with the constrained double Schechter function
  discussed in Section 6. Given the small statistical errors associated with the local KLF dataset shown in Fig. \ref{KLF_local}, none of the Schechter-function
fits are formally acceptable. As a result, the parameter uncertainties quoted in the table have been calculated after inflating the errors to enforce $\chi^{2}_{\nu}=1$.}
\begin{tabular}{ | c | c | c | c | c | c | c | c |}
\hline
\hline
Schechter fit & $\log (\phi_{1}^{*}/\rm{Mpc}^{-3})$ &  $M_{K}^{*}$ & $\alpha_{1}$ & $\log (\phi_{2}^{*}/\rm{Mpc}^{-3})$ & $\alpha_{2}$ & $\chi^{2}$ & $\chi^{2}_{\nu}$ \\ 
\hline
single  & $-$2.29 $\pm$0.02 & $-$22.35 $\pm$ 0.02 & $-$0.90 $\pm$ 0.02 &       &       & 183.6 & 3.9 \\
double & $-$2.26 $\pm$0.02 & $-$22.29 $\pm$ 0.03 & $-$0.80 $\pm$ 0.04 & $-$4.79 $\pm$ 0.53 & $-$2.35 $\pm$ 0.30 & 102.9 & 2.3 \\
double & $-$2.28 $\pm$0.02 & $-$22.16 $\pm$ 0.02 & $-$0.50 (fixed)         & $-$3.13 $\pm$ 0.03 & $-$1.50 (fixed) & 161.2 & 3.4 \\
\hline
\hline
\end{tabular}
\centering
\label{tab:local}
\end{table*}

\begin{figure*}
\centering
\includegraphics[trim = 30mm 135mm 25mm 35mm, clip]{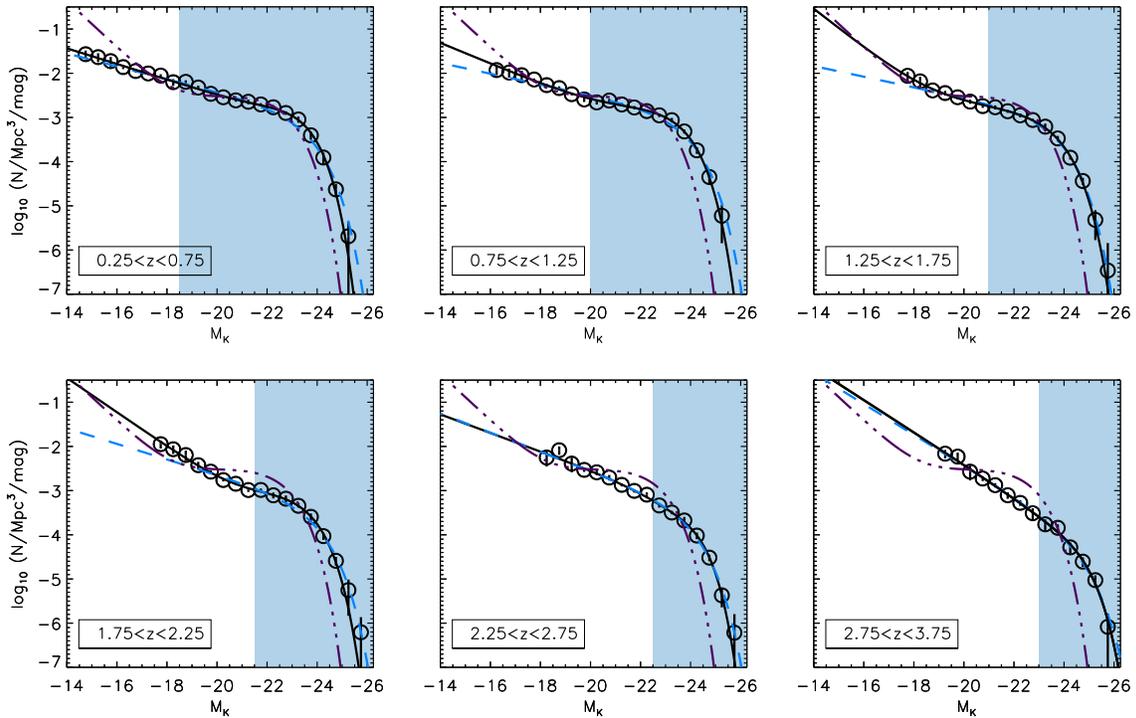}
\caption{The evolving KLF dataset together with the best-fitting
  single and double Schechter functions. In each redshift bin the black solid circles are the number
  densities of the combined UltraVISTA+CANDELS+HUDF dataset, and the
  solid black and dashed blue lines are the best-fitting double and single
  Schechter functions respectively. For reference, in each panel we also show our best-fitting double
Schechter-function fit to the local KLF dataset as the dot-dashed
purple line. The blue shaded areas
highlight the luminosity range where the data are dominated by the ground based UltraVISTA imaging.}
\label{lum_func_fits}
\end{figure*}

\begin{table*}
  \caption{The binned ($1/V_{\rm max}$) measurements of the KLF within the six redshift bins shown in Fig. 5. Column one lists the absolute $K$-band luminosity bins and
    columns $2-7$ list the individual values of $\phi_{k}$ and their corresponding uncertainties. The values of $\phi_{k}$ and their uncertainties are quoted in units of $10^{-4}$mag$^{-1}$Mpc$^{-3}$.}
\begin{tabular}{ | c | c | c | c | c | c | c |}
\hline
\hline
               & $0.25\leq z \leq0.75$ & $0.75\leq z \leq1.25$ & $1.25\leq z \leq1.75$ & $1.75\leq z \leq2.25$ & $2.25\leq z \leq2.75$ & $2.75\leq z \leq3.75$ \\ 
 & & & & & & \\
$M_{K}$ & $\phi_{k}$ & $\phi_{k}$ & $\phi_{k}$ & $\phi_{k}$ & $\phi_{k}$ &  $\phi_{k} $\\
\hline
$-$14.75 & 269.8 $\pm$ 56.8 &   &   &   &   &    \\
$-$15.25 & 231.6 $\pm$ 52.4 &    &   &   &    &    \\
$-$15.75 & 187.0 $\pm$ 46.3 &   &    &    &    &    \\
$-$16.25 & 137.7 $\pm$ 13.6 & 118.0 $\pm$ 27.2 &   &  &  &   \\ 
$-$16.75 & 112.3 $\pm$ \phantom{0}8.4 & 103.7 $\pm$ 23.9 &  &  &  &  \\
$-$17.25 & \phantom{0}98.9 $\pm$ \phantom{0}8.6 &  \phantom{0}90.5 $\pm$ 23.8 &   &   &   &  \\ 
$-$17.75 & \phantom{0}88.5 $\pm$ \phantom{0}6.7 &  \phantom{0}72.3 $\pm$ \phantom{0}4.4 &  \phantom{0}87.4 $\pm$ \phantom{0}19.3 & 112.4 $\pm$ \phantom{0}22.8 &  & \\ 
$-$18.25 & \phantom{0}62.8 $\pm$ 10.9 &  \phantom{0}53.8 $\pm$ \phantom{0}3.9 &  \phantom{0}65.6 $\pm$ \phantom{0}17.8 & \phantom{0}86.3 $\pm$ \phantom{0}18.7 &\phantom{0}56.0 $\pm$ \phantom{0}21.0 & \\ 
$-$18.75 & \phantom{0}63.9 $\pm$ \phantom{0}5.3 &  \phantom{0}46.2 $\pm$ \phantom{0}4.4 & \phantom{0}40.9 $\pm$ \phantom{00}5.0 & \phantom{0}63.8 $\pm$ \phantom{0}16.6 &\phantom{0}81.9 $\pm$ \phantom{0}17.6 &  \\ 
$-$19.25 & \phantom{0}47.5 $\pm$ \phantom{0}4.7 &  \phantom{0}33.6 $\pm$ \phantom{0}2.1 &  \phantom{0}35.6 $\pm$ \phantom{00}4.3 & \phantom{0}37.6 $\pm$ \phantom{00}7.3  &\phantom{0}40.9 $\pm$ \phantom{0}15.0 &\phantom{0}68.4 $\pm$ \phantom{0}14.8 \\ 
$-$19.75 & \phantom{0}34.2 $\pm$ \phantom{0}2.2 &  \phantom{0}25.6 $\pm$ \phantom{0}3.1 &  \phantom{0}28.6 $\pm$ \phantom{00}2.8 & \phantom{0}26.8 $\pm$ \phantom{00}1.7 &\phantom{0}29.4 $\pm$ \phantom{00}2.9 &\phantom{0}58.7 $\pm$ \phantom{0}13.3 \\ 
$-$20.25 & \phantom{0}28.6 $\pm$ \phantom{0}2.1 &  \phantom{0}21.9 $\pm$ \phantom{0}2.3 &  \phantom{0}22.8 $\pm$ \phantom{00}2.5 & \phantom{0}17.3 $\pm$ \phantom{00}1.2 &\phantom{0}26.0 $\pm$ \phantom{00}3.7 &\phantom{0}26.8 $\pm$ \phantom{00}9.9 \\ 
$-$20.75 & \phantom{0}24.4 $\pm$ \phantom{0}1.7 &  \phantom{0}24.1 $\pm$ \phantom{0}2.2 & \phantom{0}18.1 $\pm$ \phantom{00}1.6 & \phantom{0}14.3 $\pm$ \phantom{00}1.2 &\phantom{0}19.7 $\pm$ \phantom{00}2.8 &\phantom{0}18.7 $\pm$ \phantom{00}2.3 \\
$-$21.25 & \phantom{0}22.6 $\pm$ \phantom{0}2.7 &  \phantom{0}19.7 $\pm$ \phantom{0}1.6 & \phantom{0}17.0 $\pm$ \phantom{00}1.8 & \phantom{0}10.3 $\pm$ \phantom{00}1.0 &\phantom{0}13.5 $\pm$ \phantom{00}1.1 &\phantom{0}13.3 $\pm$ \phantom{00}1.5 \\
$-$21.75 & \phantom{0}19.7 $\pm$ \phantom{0}1.3 &  \phantom{0}16.8 $\pm$ \phantom{0}1.4 & \phantom{0}13.8 $\pm$ \phantom{00}1.1 & \phantom{0}10.3 $\pm$ \phantom{00}0.8 &\phantom{00}9.9 $\pm$ \phantom{00}0.9 &\phantom{00}7.8 $\pm$ \phantom{00}1.3 \\
$-$22.25 & \phantom{0}17.0 $\pm$ \phantom{0}1.9 &  \phantom{0}13.9 $\pm$ \phantom{0}1.2 & \phantom{0}11.2 $\pm$ \phantom{00}1.2 & \phantom{00}7.8 $\pm$ \phantom{00}1.2 &\phantom{00}8.1 $\pm$ \phantom{00}1.1 & \phantom{00}5.3 $\pm$ \phantom{00}0.8 \\
$-$22.75 & \phantom{0}12.6 $\pm$ \phantom{0}1.2 &  \phantom{0}11.2 $\pm$ \phantom{0}1.0 & \phantom{00}8.8 $\pm$ \phantom{00}0.6 & \phantom{00}6.6 $\pm$ \phantom{00}1.2 &\phantom{00}4.6 $\pm$ \phantom{00}0.5 & \phantom{00}3.1 $\pm$ \phantom{00}0.9 \\ 
$-$23.25 & \phantom{00}9.0 $\pm$ \phantom{0}1.4 & \phantom{00}8.7 $\pm$ \phantom{0}1.0 & \phantom{00}6.2 $\pm$ \phantom{00}1.1 & \phantom{00}4.5 $\pm$ \phantom{00}0.4 &\phantom{00}3.2 $\pm$ \phantom{00}0.3 & \phantom{00}1.7 $\pm$ \phantom{00}0.5 \\ 
$-$23.75 & \phantom{00}3.9 $\pm$ \phantom{0}0.8 & \phantom{00}4.8 $\pm$ \phantom{0}0.5 & \phantom{00}3.4 $\pm$ \phantom{00}0.5 & \phantom{00}2.6 $\pm$ \phantom{00}0.5 & \phantom{00}2.1 $\pm$ \phantom{00}0.4 & \phantom{00}1.4 $\pm$ \phantom{00}0.3 \\
$-$24.25 & \phantom{00}1.2 $\pm$ \phantom{0}0.4 & \phantom{00}1.8 $\pm$ \phantom{0}0.3 & \phantom{00}1.2 $\pm$ \phantom{00}0.1 & \phantom{0}0.93 $\pm$ \phantom{00}0.2 & \phantom{0}0.96 $\pm$ \phantom{00}0.2 & \phantom{00}0.5 $\pm$ \phantom{00}0.1 \\
$-$24.75 & \phantom{0}0.24 $\pm$ 0.07 & \phantom{0}0.45 $\pm$ 0.13 & \phantom{0}0.36 $\pm$ \phantom{0}0.07 & \phantom{0}0.26 $\pm$ \phantom{0}0.06 & \phantom{0}0.30 $\pm$ \phantom{0}0.05 & \phantom{0}0.25 $\pm$ \phantom{0}0.06 \\
$-$25.25 & \phantom{0}0.02 $\pm$ 0.02 & \phantom{0}0.06 $\pm$ 0.05 &   0.048 $\pm$  0.031 & 0.056 $\pm$ 0.041 & 0.043 $\pm$ 0.020 & 0.094 $\pm$ 0.027 \\
$-$25.75 &                               &                                                            & 0.003 $\pm$ 0.010 & 0.006 $\pm$ 0.074 & 0.006 $\pm$ 0.010 & 0.008 $\pm$ 0.008 \\ 
\hline
\hline
\end{tabular}
\centering
\label{phi_tab}
\end{table*}

\section{The evolving $\bmath{K-}$band luminosity function}
\label{sec:results}
Our new determination of the evolving KLF over the redshift interval
$0.25 \leq z \leq 3.75$ is provided in Table \ref{phi_tab} and plotted
in Fig. \ref{lum_func_fits}. In each panel of Fig. \ref{lum_func_fits}
the best-fitting single Schechter
function is shown as the dashed blue line and the best-fitting double
Schechter function is shown as the solid black line. For reference, in
each panel we also show our best-fitting double Schechter function to
the local KLF as the dot-dashed purple line.

The light blue shaded region in each panel of Fig. \ref{lum_func_fits} indicates the absolute magnitude range over which the final
galaxy sample is dominated by the ground-based data from UltraVISTA,
with the deep {\it HST} imaging from CANDELS and the HUDF dominating
at fainter magnitudes.

\begin{table*}
\caption{The best-fitting single and double Schechter parameters based
  on fitting the KLF data shown in Fig. \ref{lum_func_fits} and Table \ref{phi_tab}.
The first column lists the redshift
  bin and columns $2-6$ list the best-fitting Schechter function
  parameters and their corresponding uncertainties. Columns seven and
eight list the corresponding values of $\chi^{2}$ and reduced $\chi^{2}_{\nu}$ respectively.}
\begin{tabular}{ | c | c | c | c | c | c | c | c |}
\hline
\hline
Redshift Range & $\log (\phi_{1}^{*}/\rm{Mpc}^{-3})$ &  $M_{K}^{*}$ &
$\alpha_{1}$ & $\log (\phi_{2}^{*}/\rm{Mpc}^{-3})$ & $\alpha_{2}$ & $\chi^{2}$ & $\chi^{2}_{\nu}$ \\ 
\hline
0.25$<z<$0.75 & $-$2.89 $\pm$ 0.06 & $-$23.50 $\pm$ 0.12 & $-$1.36 $\pm$ 0.02 &       &       & 45.3 & 2.5 \\
0.75$<z<$1.25 & $-$2.92 $\pm$ 0.07 & $-$23.71 $\pm$ 0.13 & $-$1.31 $\pm$ 0.03 &       &       & 44.8 & 3.0 \\
1.25$<z<$1.75 & $-$2.95 $\pm$ 0.05 & $-$23.56 $\pm$ 0.07 & $-$1.30 $\pm$ 0.03 &       &       & 12.1 & 0.9 \\
1.75$<z<$2.25 & $-$3.28 $\pm$ 0.12 & $-$23.83 $\pm$ 0.18 & $-$1.44 $\pm$ 0.07 &       &       & 39.1 & 3.0 \\
2.25$<z<$2.75 & $-$3.36 $\pm$ 0.06 & $-$23.89 $\pm$ 0.10 & $-$1.54 $\pm$ 0.04 &       &       & \phantom{0}7.7 & 0.6 \\
2.75$<z<$3.75 & $-$3.96 $\pm$ 0.16 & $-$24.50 $\pm$ 0.20 & $-$1.87 $\pm$ 0.15 &       &       & \phantom{0}4.8 & 0.5 \\

\hline
0.25$<z<$0.75 & $-$2.59 $\pm$ 0.06 & $-$22.77 $\pm$ 0.16 & $-$0.28 $\pm$ 0.29 & $-$2.95 $\pm$ 0.10 & $-$1.44 $\pm$ 0.04 & 11.7 & 0.7 \\
0.75$<z<$1.25 & $-$2.65 $\pm$ 0.06 & $-$23.16 $\pm$ 0.15 & $-$0.72 $\pm$ 0.23 & $-$3.44 $\pm$ 0.29 & $-$1.59 $\pm$ 0.11 & \phantom{0}8.4 & 0.7 \\
1.25$<z<$1.75 & $-$2.82 $\pm$ 0.07 & $-$23.36 $\pm$ 0.13 & $-$1.11 $\pm$ 0.16 & $-$4.67 $\pm$ 1.38 & $-$2.11 $\pm$ 0.58 & \phantom{0}2.9 & 0.3 \\
1.75$<z<$2.25 & $-$2.88 $\pm$ 0.07 & $-$23.17 $\pm$ 0.16 & $-$0.77 $\pm$ 0.23 & $-$3.97 $\pm$ 0.33 & $-$1.97 $\pm$ 0.15 & \phantom{0}6.7 & 0.6 \\
2.25$<z<$2.75 & $-$3.49 $\pm$ 0.18 & $-$23.18 $\pm$ 0.30 & \phantom{$-$}0.26 $\pm$ 0.69 & $-$3.17 $\pm$ 0.15 & $-$1.53 $\pm$ 0.08      & \phantom{0}5.7 & 0.6 \\
2.75$<z<$3.75 & $-$3.91 $\pm$ 0.36 & $-$23.92 $\pm$ 0.66 & $-$0.59 $\pm$ 1.46 & $-$3.81 $\pm$ 0.45 & $-$1.91 $\pm$ 0.15 & \phantom{0}3.7 & 0.5 \\
\hline
\hline
\end{tabular}
\centering
\label{tab:LF_params}
\end{table*}

\begin{figure*}
\centering
\includegraphics[trim = 25mm 135mm 25mm 25mm, clip]{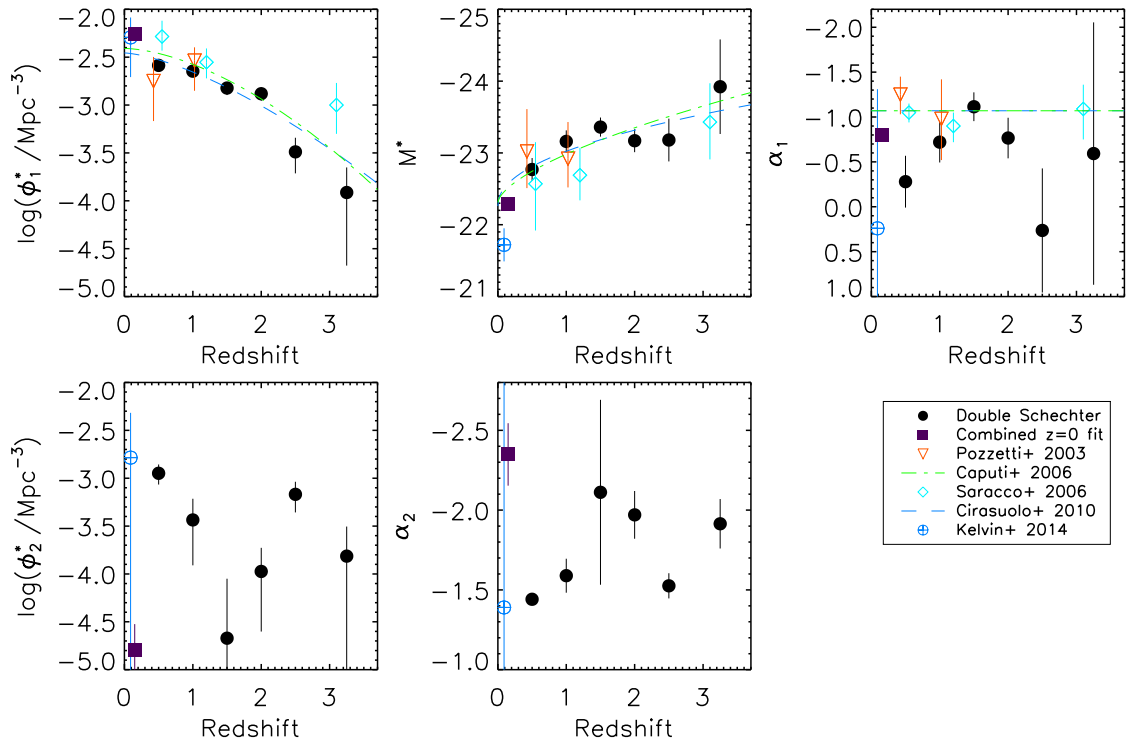}
\caption{The redshift evolution of the parameters describing the double Schechter
  function fits to the KLF dataset. The top three panels show the
  evolution of the parameters of the Schechter component which
  describes the bright end of the KLF. The two bottom panels show the
  evolution of the Schechter component which describes the up-turn in
  galaxy number density seen at faint magnitudes. In addition to the
  results for the six redshift bins shown in Fig. \ref{lum_func_fits},
  we also plot the parameters derived from the local KLF study of
  \citet{Kelv14} and our own double Schechter-function fit to
  the combined local KLF dataset shown in Fig. \ref{KLF_local}. The
  green and blue lines plotted in the upper panels are the fits to the
  evolving (single Schechter) parameters derived by Caputi et
  al. (2006) and \citet{Cira10} respectively (both studies fixed the faint-end slope at $\alpha=-1.07$).} 
\label{lum_func_double_params}
\end{figure*}

\begin{figure*}
\centering
\includegraphics[trim = 30mm 135mm 25mm 35mm, clip]{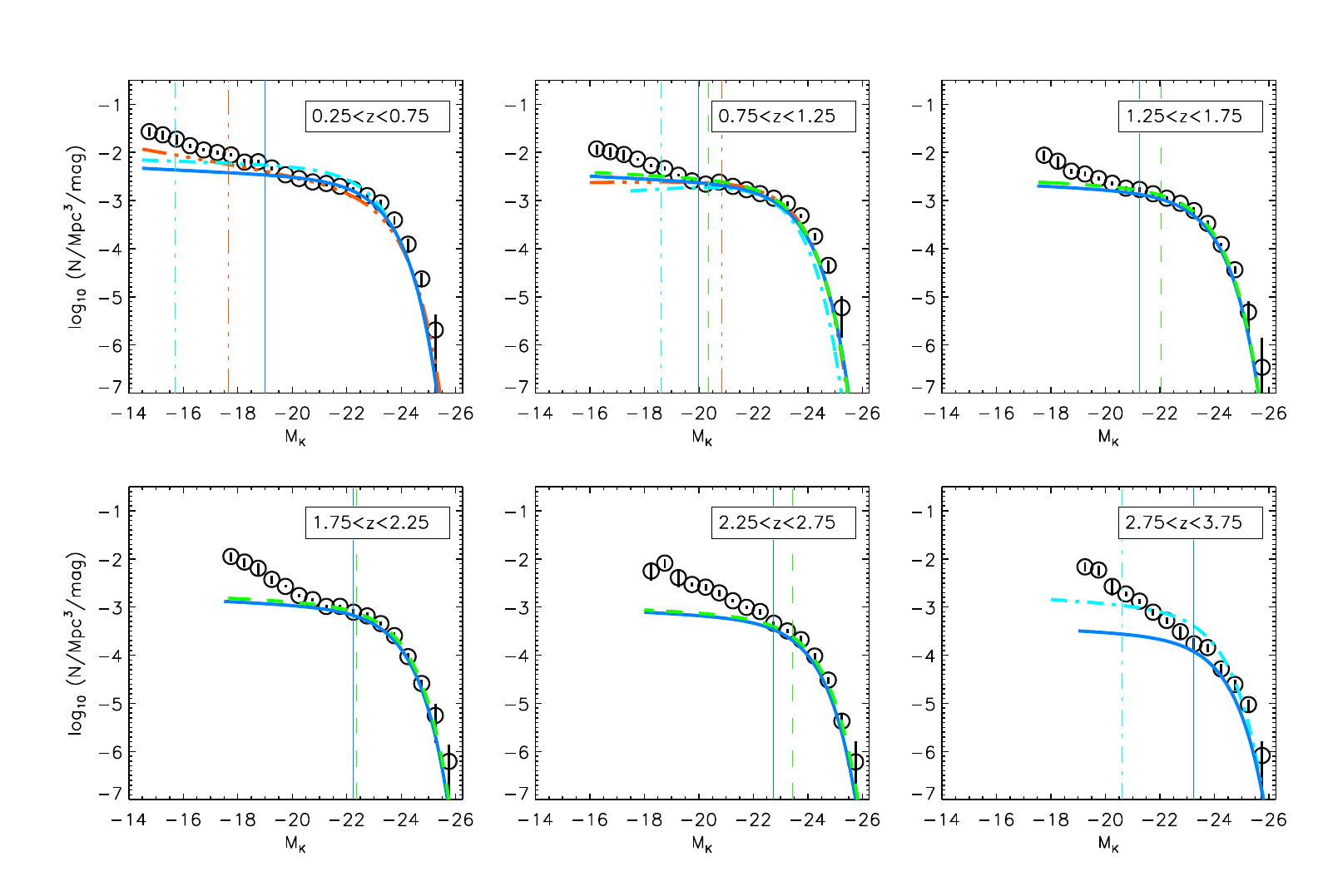}
\caption{A comparison between our new measurement of the evolving KLF and the Schechter
  function fits derived by previous observational studies. The red triple-dot-dashed lines show the
  Schechter-function fits from \citet{Pozz03}, the light blue dot-dashed
  lines are the fits from \citet{Sara06}, the green dashed lines are
  the fits from \citet{Capu06} and the blue solid lines are the
  fits from \citet{Cira10}. The corresponding vertical lines are the magnitude limits of the previous studies.}
\label{lum_func_lit}
\end{figure*}

\subsection{Single versus Double Schechter fits}
Although it is now established that the local KLF cannot be well
reproduced by a single Schechter function (e.g. \citealt{Smit09};
\citealt{Kelv14}), previous studies of the evolving KLF have not possessed the combination of wide area and depth necessary to
accurately determine its functional form at $z\geq 1$
(e.g. \citealt{Capu06}; \citealt{Cira10}). The galaxy sample
assembled for this study allows us to test the functional form of the
evolving KLF at $z\geq 1$ for the first time.

The best-fitting parameters and the corresponding minimum $\chi^{2}$
values for the Schechter-function fits shown in Fig. \ref{lum_func_fits}
are provided in Table \ref{tab:LF_params}. As expected, the double 
Schechter function provides a better fit to the KLF data in all six redshift
bins. However, it is not the case that the double Schechter fit can be
statistically preferred to the single Schechter fit in all cases.
Fortunately, given that the single Schechter-function fits form a {\it
  nested} sub-set of the double Schechter-function fits,  it is 
straightforward to decide whether the double fit is statistically preferred. 
In this scenario, we expect the $\Delta
\chi^{2}$ between the best-fitting models (i.e. $\Delta \chi^{2} = \chi^{2}_{\rm{single}} - \chi^{2}_{\rm{double}}$) to follow a $\chi^{2}$
distribution with two degrees of freedom, since the double Schechter
has two more free parameters than the single Schechter function. 
Consequently, if the double fit is to be preferred over the
single fit at the $99\%$ confidence level, we require a value of $\Delta
\chi^{2}\geq 9.2$ between the two competing model fits.

It can be seen from the information presented in Table
\ref{tab:LF_params} that the double Schechter-function fit is formally
preferred to the single Schechter-function fit at $\geq 99\%$
confidence within the first four redshift bins. Within the final two redshift
bins the double Schechter function is not
statistically preferred over the single fit. This conclusion agrees
well with a visual inspection of Fig. \ref{lum_func_fits}, which also indicates
that the extra freedom provided by the double Schechter function is
not actually required to describe the data in the two highest redshift bins.

Overall it is clear that our new UltraVISTA+CANDELS+HUDF dataset
indicates that the KLF has a double Schechter form out to redshifts of
$z\simeq 2$, but that a double Schechter function is not formally required to
describe the KLF data at $2.25\leq z \leq 3.75$. However, it is not immediately clear whether the apparent change in functional form at $z\simeq 2$ indicates a genuine transition or, alternatively, it simply
reflects a combination of smoothing of the intrinsic KLF features due to photometric redshift uncertainties and the inevitable bias towards
deriving steep faint-end slopes when dealing with a reduction in the
available dynamic range in luminosity (c.f. \citealt{Pars16}).
This issue is discussed further in Section 6. 

\subsection{Evolving Schechter function parameters}
The redshift evolution of the best-fitting double Schechter-function
parameters is shown in Fig. \ref{lum_func_double_params}.  The panels
in the top row show the evolution of the three parameters which
describe the Schechter component that dominates the bright end of the
KLF, whereas the bottom panels show the evolution of the normalisation and slope of
the component that dominates the faint end of the KLF.

The top-left panel shows a steady decrease in the value of 
$\log(\phi_{1}^{\star}/{\rm Mpc^{-3}})$, from a local value of $\simeq -2.3$ to
a value of $\simeq -4.0$ by $z\simeq 3.5$. Likewise, it can be seen that
$M_{K}^{\star}$ also shows a relatively smooth evolution with
redshift, changing from $\simeq -22.3$ locally to $\simeq -23.8$ by $z\simeq 3.5$.
In contrast, the value of $\alpha_{1}$ shows no real evolutionary trend, with a mean (median) value of $\langle \alpha_{1}\rangle=-0.54\pm0.18 (-0.66)$.
Likewise, the bottom panels of Fig. \ref{lum_func_double_params} suggests that
neither $\alpha_{2}$ or $\phi_{2}^{\star}$ show any convincing trend with
redshift, with mean (median) values of $\langle \alpha_{2} \rangle=-1.76\pm0.10
 (-1.75)$ and $\langle \log(\phi_{2}^{\star}/{\rm  Mpc^{-3}}\rangle=-3.67\pm0.23 (-3.63)$ respectively.

To first order,  the parameters shown in the top row are expected to
mimic those that would be obtained by fitting a single Schechter
function to only the bright end of our KLF data set (i.e. $M_{K}\leq -21$). 
To illustrate this point, in the top row of Fig. \ref{lum_func_double_params} we plot the 
parameter values derived by four previous studies of the evolving KLF,
based on fitting single Schechter functions to data with a lower dynamic range in $K$-band luminosity. 
Within the errors, it can be seen that the parameters derived by previous
literature studies, using single Schechter-function fits to only the
bright end of the evolving KLF, agree with the ($M_{K}^{\star},
\phi_{1}^{\star}$) parameters derived here by fitting a double
Schechter function over a much greater dynamic range in $K$-band luminosity.

In summary, the results plotted in Fig. \ref{lum_func_double_params} suggest that it may be possible to describe the evolution of the KLF using a double Schechter
function, in which the bright-end component evolves smoothly with redshift while the
faint-end component remains approximately constant. This prospect is pursued further in Section 6.

\subsection{Comparison to previous observational results}
To explore the agreement with previous literature results further, in
Fig. \ref{lum_func_lit} we compare our new KLF dataset with 
the best-fitting single Schechter functions derived by \citet{Pozz03}, \citet{Capu06}, 
\citet{Sara06} and \citet{Cira10}.

As can readily be seen from Fig. \ref{lum_func_lit}, the single
Schechter-function fits derived by the four previous literature
studies continue to provide a good description of our new KLF dataset,
at least down to the magnitude limits explored by the previous
studies.  In the first two redshift bins there is some evidence that our
bright-end data-points are somewhat brighter than the literature
Schechter-function fits. However, this is expected given our
revised treatment for correcting to total magnitudes which accounts for extended light at large radii.

The comparison shown in Fig. \ref{lum_func_lit} clearly demonstrates that the relatively flat faint-end slopes (i.e. $\alpha\simeq -1$) derived
(or assumed) by previous studies completely fail to describe the
up-turn in the number density of galaxies at fainter magnitudes revealed by our new UltraVISTA+CANDELS+HUDF dataset.
This graphically re-enforces the conclusion that a double Schechter
function is necessary to simultaneously match the steep decrease
in number density at $M_{K}<M_{K}^{\star}$ and the up-turn seen at fainter magnitudes.

\subsection{Comparison to simulation results}
\label{sec:LF_sims_comp}

\begin{figure*}
\centering
\includegraphics[trim = 30mm 135mm 25mm 35mm, clip]{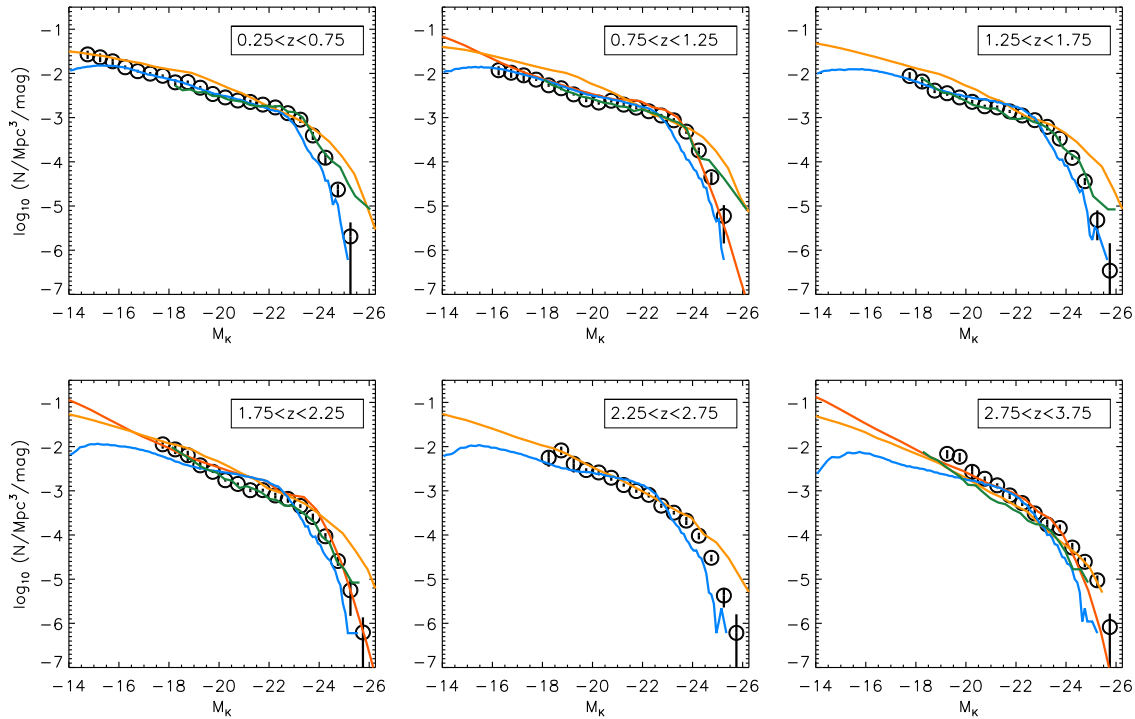}
\caption{
  A comparison between our new measurement of the evolving KLF and the predictions of four recent galaxy-evolution models.
  The yellow line shows the predictions of the Illustris hydro-dynamical simulation  \citep{Gene14}, the dark green line shows the predictions of the \textsc{mufasa} hydro-dynamical simulation \citep{Dave16},
  the red line shows the predictions of the \citet{Henr15} semi-analytic model and the blue line shows the predictions of the  \citet{Gonz14} semi-analytic model. Predictions from the \citet{Henr15} and \citet{Dave16} models are not presented in all redshift bins.}
\label{lum_func_sims}
\end{figure*}

\label{lf_evo}
\begin{figure*}
\centering
\includegraphics[trim = 35mm 130mm 30mm 65mm, clip]{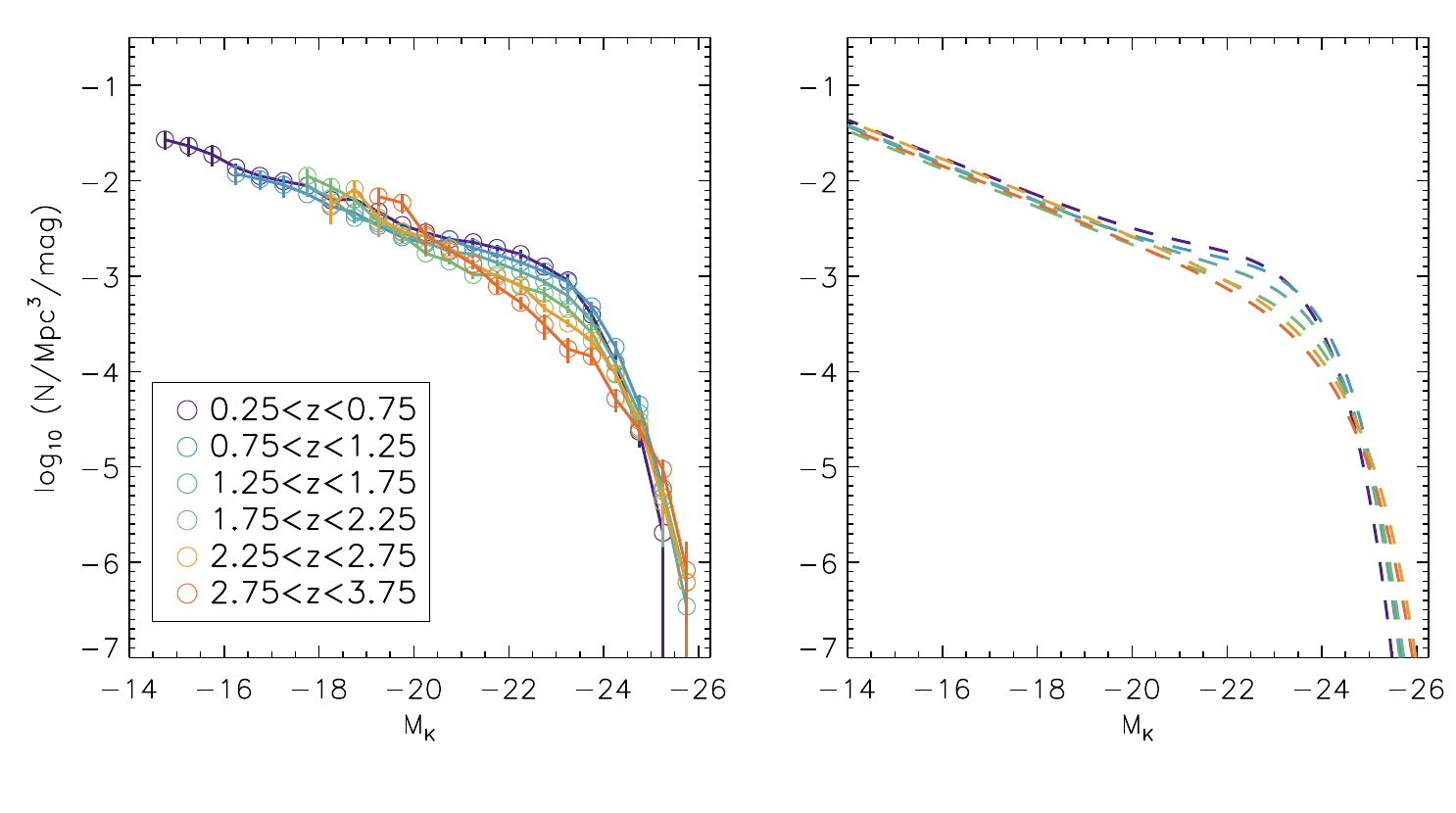}
\caption{The left-hand panel shows an overlay of the KLF data from all six redshift bins.
  It can be seen immediately that the KLF data is consistent
  with having the same faint-end slope at $M_{K}\geq-20$. Moreover, it is clear that there appears to be very little evolution
  in the number density of galaxies brighter than  $M_{K}\simeq-24.5$. Consequently, over the redshift interval $0.25\leq z \leq 0.75$ it appears that the evolution
of the KLF consists largely of a relatively smooth build-up in the number density of intermediate luminosity galaxies,
with absolute magnitudes in the range $-20\leq M_{K} \leq -24$. The right-hand panel shows the reproduction of the observed KLF evolution based on our simplified, 3-parameter prescription (see Section \ref{sec:simple_prescrip} for a discussion).}
\label{lf_overlay}
\end{figure*}

In Fig.  \ref{lum_func_sims} we compare our new KLF dataset to
the predictions of four recent galaxy evolution
simulations, two of which are semi-analytic in nature and two of which are hydro-dynamical.
The first hydro-dynamical model is Illustris (yellow line; \citealt{Gene14}), which is an
N-body/hydro-dynamical simulation in which the physics governing
galaxy formation and evolution is tuned to match the local galaxy
stellar mass function (GSMF)
and the evolution of the cosmic star-formation rate density. The second hydro-dynamical model is the
recent \textsc{mufasa} simulation (dark green line; \citealt{Dave16}), which
employs an empirical prescription for quenching based on halo
mass. 

The red line in Fig.  \ref{lum_func_sims} shows the predictions
of the \cite{Henr15} semi-analytic galaxy evolution model. This
simulation is based on the 
Munich galaxy formation models, but includes updates which
match observations of the passive fraction of galaxies in the redshift
range $0<z<3$, as well as the evolution of the GSMF. Finally, the blue
line in Fig.  \ref{lum_func_sims} shows the predictions of the \citet{Gonz14} semi-analytic model, which is a recent update of  \textsc{galform} (\citealt{Cole00}).
Importantly, the absolute $K$-band magnitudes predicted by all four models shown in Fig. \ref{lum_func_sims} are based on Bruzual \& Charlot stellar population models, and
should therefore be directly comparable to our KLF dataset.

In the first three redshifts bins shown in Fig. \ref{lum_func_sims}, the predictions
from  the \citet{Gonz14} model do a good job of reproducing the observed
normalisation and faint-end slope of the KLF. However, at these redshifts
there is a clear tendency to under-predict the number density of bright
galaxies ($M_{K}\leq -23$), which appears to be the result of
predicting a value of $M_{K}^{\star}$ that is $\simeq 0.5-1$
mag fainter than observed. In the three higher redshift bins
the difference between  predicted and observed $M_{K}^{\star}$ continues and
the KLF predicted by the \citet{Gonz14} model also displays a faint-end slope that is
somewhat shallower than is observed.

The predictions of the \citet{Henr15} semi-analytic model do a good job of
reproducing the normalisation and faint-end slope of the observed KLF over the redshift range $0.75 \leq z \leq 3.75$.
In particular, it is noticeable that the \citet{Henr15} model is able to accurately reproduce the bright end of the observed KLF at $z\simeq 1$ and $z\simeq 2$.
In the highest redshift bin at $z\simeq 3.25$, the \citet{Henr15} model produces the best overall match to the observed data, although it does under-predict
the number density of the very brightest galaxies.

The \textsc{mufasa} simulation does an excellent job of matching the normalisation,
faint-end slope and break in the observed KLF at $z\leq 2.25$, displaying only a slight
tendency to over-produce the very brightest galaxies  at $z\leq 1.75$. Interestingly, it appears that the \textsc{mufasa} model can
reproduce the inflection point observed in the KLF between the bright end and the up-turn seen at fainter magnitudes.
In the highest redshift bin at $z\simeq 3.25$ the \textsc{mufasa} model mimics the shape of the observed KLF very well, but under-predicts the
observed number densities by a constant factor of $\simeq 2$.

Finally, the Illustris model does a good job of matching the observed number
densities around $M_{K}^{\star}$ at all redshifts, but consistently over-predicts the
observed number densities at fainter and brighter magnitudes.  Although the Illustris predictions shown in Fig. \ref{lum_func_sims} do not include dust reddening,
the tendency to over-predict the number density of galaxies at the extreme ends of the KLF is entirely
consistent with comparisons between the Illustris simulation and
observations of the evolving GSMF (\citealt{Gene14}).

Overall, the comparison between the latest model predictions and our new determination of the evolving KLF are encouraging, with the \cite{Henr15}
semi-analytic and \textsc{mufasa} (\citealt{Dave16})
hydro-dynamical simulations doing a particularly good job of
reproducing the observed data. In particular, although substantial differences
remain in detail, particularly at the bright end, all of the models agree reasonably well on the basic shape/slope of the KLF at magnitudes fainter than the break.

The last detailed comparison between observations of the evolving KLF and simulation predictions was performed by \citet{Cira10}. At that time, simulation
results appeared to significantly over-predict the number density of faint galaxies, although the limited dynamic range of the data available to \citet{Cira10} only allowed the
comparison to be made at magnitudes brighter than $M_{K} \simeq -20, M_{K} \simeq -22$ and $ M_{K} \simeq -23$ at $z=1, 2, \&\, 3$ respectively.

The improvements in the quality of observational data
over the last five years mean that it is now possible to perform this comparison to much fainter magnitudes over the full redshift range. Although the simulation results presented in Fig. \ref{lum_func_sims} have typically been tuned to reproduce the stellar mass function (and therefore the KLF)
at low redshift, it is encouraging that they continue to agree reasonably well, at least qualitatively, with the observed KLF over a broad range in redshift.

\begin{figure*}
\centering
\includegraphics[trim = 35mm 140mm 55mm 20mm, clip, width=16cm]{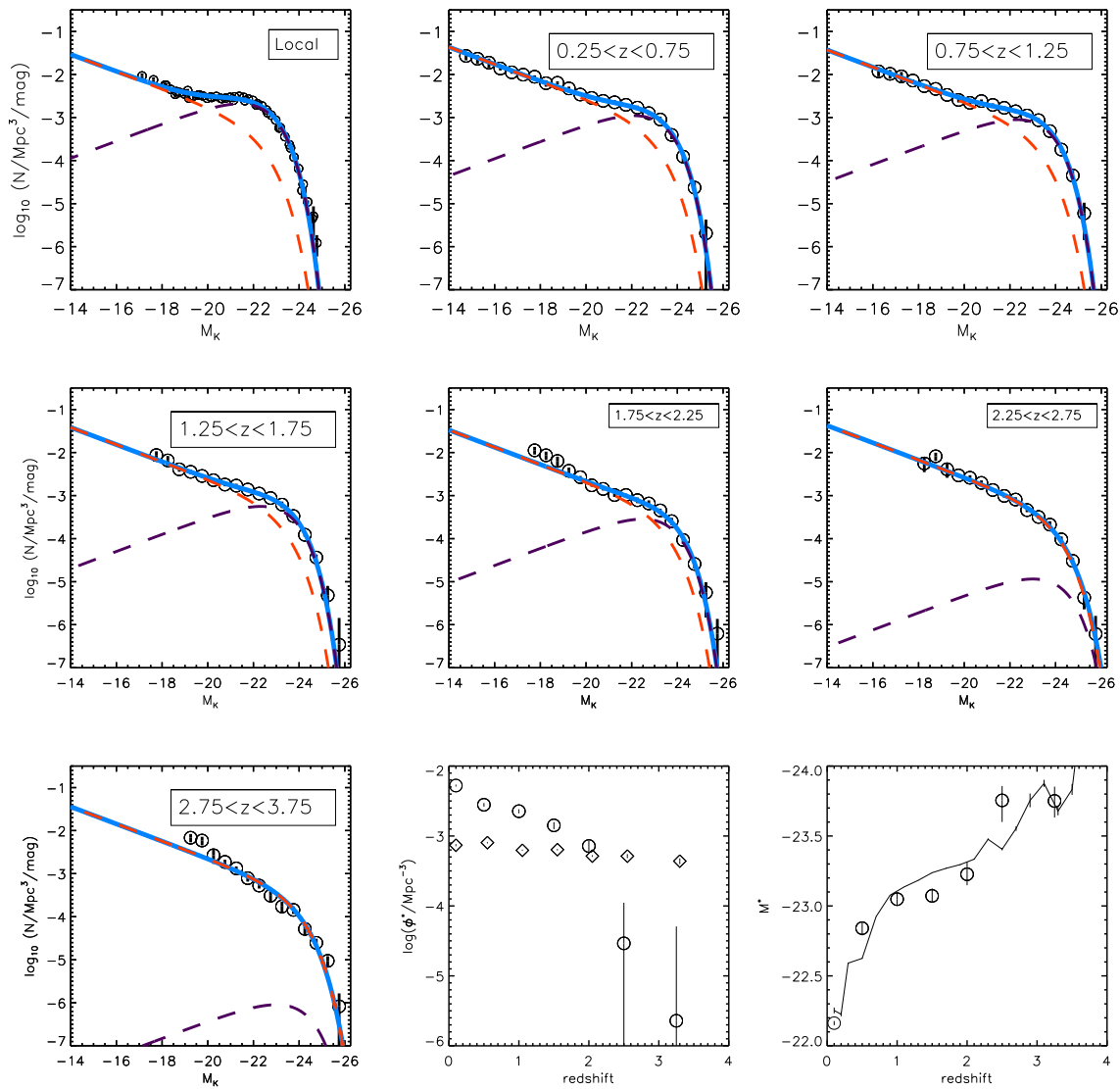}
\caption{A comparison between our new measurement of the evolving KLF and the best-fitting double Schechter
  function with a shared value of $M_{K}^{\star}$ and fixed faint-end slopes of $\alpha_{1}=-0.5$ and $\alpha_{2}=-1.5$. In each panel the solid blue line shows the sum of
  the Schechter function component that dominates at faint magnitudes (red dashed line) and the Schechter function component which dominates
  the bright end of the KLF at low redshifts (purple dashed line). The top-left panel shows the best-fit to the local KLF dataset under this simplified, 3-parameter prescription.
  The two panels at the bottom right show the redshift evolution of the three free parameters: $\phi_{1}^{\star}$ (open circles), $\phi_{2}^{\star}$ (diamonds) and $M_{K}^{\star}$. The solid line in the bottom-right panel
  shows the expected evolution of $M_{K}^{\star}$ if it is assumed to correspond to a constant stellar mass of $M_{\star}\simeq 5\times 10^{10}\Msun$ at all redshifts (see text for a discussion).
}

\label{lf_constrained}
\end{figure*}

\section{A simple prescription for the evolution of the KLF}
\label{sec:simple_prescrip}
The analysis of the evolving KLF in the previous section has
established two clear facts. Firstly, it is necessary to
employ a double Schechter function in order to reproduce the observed KLF at $z\leq 2$.
Secondly, the results of separately fitting a double
Schechter function to each redshift bin suggest that the evolving KLF can be described by the combination of 
a smoothly evolving bright-end component and an approximately constant faint-end component.

These conclusions are strengthened further by the left-hand panel of Fig. \ref{lf_overlay},
which shows an overlay of the KLF data from all six redshift
bins. This plot gives the impression that if sufficient dynamic range in luminosity were available, the 
KLF would be observed to have a roughly constant normalisation and faint-end slope at $M_{K}\geq -20$.
Moreover, over the redshift interval $0.25 \leq z \leq 3.75$ there
appears to be surprisingly little evolution, perhaps no evolution, in the number
density of the brightest galaxies at $M_{K}\leq -24$.
Indeed, over the redshift interval studied here, the left-hand panel of Fig. \ref{lf_overlay} strongly suggests that the evolution of the KLF
largely consists of a smooth build-up in the number density of intermediate luminosity galaxies within the absolute magnitude range $-20 \leq M_{K} \leq -24$. Similar results have been found for the UV luminosity function in \citet{Bowl15}. 

This scenario immediately suggests that it would be interesting to
explore the evolution of the KLF within the context of the
phenomenological galaxy evolution model proposed by \citet{Peng10}.
The \citet{Peng10} model describes the total GSMF at $z\leq 2$ as a double Schechter function with a shared
value of $M_{\star}\simeq 5 \times10^{10}\Msun$, the value of which is governed by the process of mass quenching of star formation.

In this model the overall double Schechter-function shape of the GSMF is produced by the combination of an approximately constant star-forming
component and a rapidly evolving quenched component. The star-forming component is described by a single Schechter function with a fixed
faint-end slope of $\alpha_{SF} \simeq -1.3$, whereas the quenched component has a double Schechter functional form, with a faint-end slope of $\alpha_{SF}$ and a
bright-end slope of $\alpha_{Q}=\alpha_{SF}+1$. The combination of star-forming and quenched components produces an overall stellar mass function with a double Schechter functional form with a faint-end slope of $\alpha_{SF}$,  a bright-end slope of $\alpha_{SF}+1$ and a shared value of $M_{\star}\simeq 5 \times10^{10}\Msun$.
Observationally, this model is known to be consistent with recent
determinations of the GSMF at $z\simeq 0$ (e.g. \citealt{Bald08}; \citealt{Bald12}) and $z\simeq 1$ (e.g. \citealt{Ilbe13}; \citealt{Tomc14}; \citealt{Mort15}).

Motivated by this, we explored re-fitting the KLF using a double
Schechter function with a shared value of $M_{K}^{\star}$, insisting that  
the slope of the component describing the faint-end of the KLF
($\alpha_{2}$) is constant with redshift and enforcing the additional constraint that $\alpha_{1}=\alpha_{2}+1$. 
The adopted value of $\alpha_{2}$ was derived from a fit to the KLF
data within the $0.25 \leq z \leq 0.75$ redshift bin which covers that largest
dynamic range in $K$-band luminosity. By stepping through a grid with
a spacing of $\Delta \alpha_{2}=\Delta \alpha_{1}=0.1$, it was determined
that ($\alpha_{1},\alpha_{2})=(-0.5,-1.5)$ provided the best-fit and these values were therefore adopted and held constant when fitting the
KLF in all six redshift bins. The results of this constrained
3-parameter fitting process are listed in Table \ref{tab:constrained} and plotted in the right-hand panel of Fig. \ref{lf_overlay}.

In Fig. \ref{lf_constrained} we compare the results of the constrained KLF fitting to the data in all six redshifts bins, illustrating how the
two components combine to provide the overall double Schechter shape. In the top-left panel of Fig. \ref{lf_constrained} we also show a constrained
fit to the local KLF dataset, highlighting that the evolving 3-parameter fit can naturally reproduce the KLF over the full $z=0$ to $z=3.75$ redshift interval.
The bottom-right panels of Fig. \ref{lf_constrained} show the redshift
evolution of three free parameters: $\phi^{\star}_{1}$,
$\phi^{\star}_{2}$ and $M_{K}^{\star}$. 

Interestingly, it can be seen that the normalisation of the Schechter component dominating the faint-end of the
KLF ($\phi^{\star}_{2}$) remains approximately constant with redshift,
decreasing by a factor of only $\simeq 2$ over a $\simeq 12$ Gyr time-frame.
In contrast, the normalisation of the Schechter component that dominates the bright end of the KLF at low redshift ($\phi^{\star}_{1}$) decreases by
an order of magnitude between $z=0$ and $z\simeq 2$, before effectively disappearing entirely at $z\geq 2.5$.
Moreover, it can be seen from the bottom right panel of Fig. \ref{lf_constrained} that the
shared value of $M_{K}^{\star}$ evolves by $\Delta M_{K}^{\star}\simeq 1.5$ magnitudes, brightening from $M_{K}^{\star}=-22.3$ locally to
$M_{K}^{\star}\simeq -23.8$ at $z\simeq 3$.

\subsection{Comparison to the stellar mass function}
Although this paper is focused on the evolution of the KLF, it is
obviously of interest to briefly consider the implications of 
our simplified description of the evolving KLF in terms of the GSMF.

Recent determinations of the local GSMF have demonstrated that it has a double Schechter functional form, with faint and bright-end slopes which
differ by approximately unity. Indeed, in their analysis of the local GSMF using SDSS data, \citet{Bald08} and \citet{Peng10} derived values of
($-0.46\pm0.09$,$-1.58\pm0.05$) and ($-0.52\pm0.04$,$-1.56\pm0.12$) for ($\alpha_{1},\alpha_{2})$ respectively. More recently, \citet{Bald12} derived values of
($-0.35\pm0.17$,$-1.47\pm0.05$) based on data from the GAMA survey. It
is clear therefore that our adopted values of ($-0.5,-1.5$) are in excellent agreement with recent
determinations of the local GSMF. Moreover, all three of the local GSMF studies mentioned above derive values of the characteristic stellar mass that are in excellent agreement: $\log({M_{\star}}/{\Msun})\simeq 10.65\pm0.01$,  $10.67\pm0.01$ and $10.66\pm0.05$ respectively (\citealt{Chab03} IMF).

Within this context, it is interesting to reconsider the evolution we
derive for the characteristic $K$-band magnitude of the KLF based on our constrained 3-parameter fits.
In the bottom-right panel of Fig. \ref{lf_constrained} the solid line shows the expected evolution of the characteristic absolute $K$-band
magnitude ($M_{K}^{\star}$), if we make the assumption that it corresponds to a constant stellar mass of $\log({M_{\star}}/{\Msun})=10.65$.
Remarkably, this comparison suggests that the evolution of the KLF between $z=0$
and $z=4$ is perfectly consistent with an evolving value of $M_{K}^{\star}$ which corresponds to
a {\it constant} stellar mass of $\simeq 5\times 10^{10}\Msun$ (\citealt{Chab03} IMF). Indeed, we note with interest that the latest determinations of the GSMF at $z\geq 4$ suggest that the characteristic stellar mass remains constant at $\simeq 5\times 10^{10}\Msun$ out to $z\simeq 5$ (\citealt{Song16}).

\begin{table*}
  \caption{The best-fitting double Schechter functions (with fixed faint-end slopes) based on fitting the KLF data shown in Fig. 5 and Table 3. The first column lists the redshift bin and columns $2-6$ list the best-fitting parameters
    and their corresponding uncertainties. Columns 7 \& 8 list the
  corresponding values of $\chi^{2}$ and reduced $\chi^{2}_{\nu}$
  respectively.}
\begin{tabular}{ | c | c | c | c | c | c | c | c |}
\hline
\hline
Redshift Range & $\log (\phi_{1}^{*}/\rm{Mpc}^{-3})$ &  $M_{K}^{*}$ &
$\alpha_{1}$ & $\log (\phi_{2}^{*}/\rm{Mpc}^{-3})$ & $\alpha_{2}$ & $\chi^{2}$ & $\chi^{2}_{\nu}$ \\ 
\hline
0.25$<z<$0.75 & $-$2.55 $^{+0.02}_{-0.03}$ & $-$22.84 $^{+0.05}_{-0.05}$ & $-$0.50 (fixed) & $-$3.10 $^{+0.01}_{-0.01}$ & $-$1.50 (fixed) &14.7 &0.8 \\[1ex]
0.75$<z<$1.25 & $-$2.65 $^{+0.03}_{-0.03}$ & $-$23.05 $^{+0.05}_{-0.05}$ & $-$0.50 (fixed) & $-$3.20 $^{+0.01}_{-0.01}$ & $-$1.50 (fixed) &\phantom{0}9.6 &0.6\\[1ex]
1.25$<z<$1.75 & $-$2.85 $^{+0.05}_{-0.05}$ & $-$23.07 $^{+0.05}_{-0.05}$ & $-$0.50 (fixed) & $-$3.19 $^{+0.01}_{-0.01}$ & $-$1.50 (fixed) &\phantom{0}6.2  &0.5\\[1ex]
1.75$<z<$2.25 & $-$3.14 $^{+0.10}_{-0.10}$ & $-$23.23 $^{+0.11}_{-0.13}$ & $-$0.50 (fixed) & $-$3.29 $^{+0.02}_{-0.02}$ & $-$1.50 (fixed) &26.9 &2.1 \\[1ex]
2.25$<z<$2.75 & $-$4.53 $^{+0.58}_{-5.46}$ & $-$23.75 $^{+0.15}_{-0.10}$ & $-$0.50 (fixed) & $-$3.29 $^{+0.03}_{-0.03}$ & $-$1.50 (fixed) &\phantom{0}8.8  &0.7\\[1ex]
2.75$<z<$3.75 & $-$5.39 $^{+1.49}_{-4.35}$ & $-$23.75 $^{+0.21}_{-0.17}$ & $-$0.50 (fixed) & $-$3.36 $^{+0.04}_{-0.04}$ & $-$1.50 (fixed) &30.9 & 3.1\\[1ex]
\hline
\hline
\end{tabular}
\centering
\label{tab:constrained}
\end{table*}

\section{Conclusions}
\label{sec:conclusions}
We have presented the results of a study of the evolving KLF, based on a new dataset
compiled from the UltraVISTA,~CANDELS and HUDF surveys.
The large dynamic range in luminosity spanned by this new dataset ($3-4$ dex
over the full redshift range) is sufficient to allow a detailed measurement of the functional form of the evolving KLF at $z\geq 1$ for the first time, and
to allow a meaningful comparison with the predictions of the latest generation of theoretical galaxy-evolution models. Our principal conclusions are as follows:

\begin{enumerate}

\item{The large dynamic range in $K$-band luminosity provided by our new dataset is sufficient to demonstrate that the faint-end slope of the KLF is steeper than typically determined by previous $z > 0.3$ studies.  When fitted with a single
  Schechter function, our data suggests that the faint-end slope lies in the range $-1.30 \leq \alpha \leq -1.54$ within the redshift interval $0.25 \leq  z \leq 2.75$. Based on a single Schechter-function fit, there is some evidence that the
  faint-end slope steepens in our final  $z\simeq 3.25$ redshift bin ($\alpha=-1.87\pm0.15$), although the reduced dynamic range in this bin means that the slope is not particularly well constrained.}

\item{A double Schechter function, with a shared value of $M_{K}^{\star}$, provides a significantly better statistical description of the KLF than a single Schechter function, at least in the redshift range $0.25 \leq z \leq 2.25$. At higher redshifts the
  available dynamic range in luminosity is insufficient to discriminate between a single and double Schechter-function fit.}

\item{Although significant differences still exist in detail, the overall shape and normalisation of the evolving KLF is found to be in reasonable agreement with the predictions of the latest generation of galaxy-evolution models.}
  
\item{Overlaying the data in all six redshift bins suggests that the evolution of the KLF is remarkably smooth. Indeed, the data suggest that the KLF is consistent with having
  a relatively constant normalisation and slope at faint magnitudes
  (i.e. $M_{K}\leq -20$) with little, or no, evolution in the number density of galaxies brighter than $M_{K}\leq -24$.
  Instead, the KLF is observed to evolve rapidly at intermediate
  magnitudes, with the number density of galaxies at $M_{K}\simeq
  -23.0$ decreasing by a factor of $\simeq 5$ between $z\simeq 0.25$
  and $z\simeq 3.75$.}

 \item{Motivated by the apparently smooth evolution, and the phenomenological model of \citet{Peng10},  we explored the possibility of describing the evolution of the
  KLF using a double Schechter function with fixed faint-end slopes,
  such that $\alpha_{1}-\alpha_{2}=1$. Based on fitting the data in
  our $0.25 \leq z \leq 0.75$ redshift bin, which spans the largest dynamic range in luminosity, the best-fitting
  values of the faint-end slopes were determined to be: $\alpha_{1}=-0.5$ and $\alpha_{2}=-1.5$, in excellent agreement with recent studies of the local GSMF. Moreover, we demonstrated that
  this ($\alpha_{1},\alpha_{2}$) combination also provides a good description of recent determinations of the local KLF.}

 \item{We find that this simple 3-parameter fit  ($M_{K}^{\star}, \phi_{1}^{\star}, \phi_{2}^{\star}$) provides a remarkably good description of the evolving KLF,
   in which the normalisation of the component dominating the
   faint-end remains approximately constant, decreasing by a factor of
   only $\simeq 2$ over the full redshift range. In contrast, the normalisation of the
   component dominating the bright end of the KLF at low redshift evolves rapidly, decreasing by an order of magnitude between $z=0$ and $z\simeq 2$ and becoming negligible at $z\geq 2.5$.
  Moreover, within this framework, the value of the characteristic $K$-band absolute magnitude evolves by $\Delta M_{K}^{\star}\simeq 1.5$ magnitudes,
  brightening  from a local value of $M_{K}^{\star}\simeq -22.3$ to
  $M_{K}^{\star}\simeq -23.8$ by $z\simeq 3$. This evolution is shown to be entirely consistent with the underlying stellar mass function
  having a constant characteristic mass of $M_{\star}\simeq 5\times 10^{10}\Msun$ at all redshifts.}

\end{enumerate}

\section*{Acknowledgments}
AM, RJM, DJM and EMQ acknowledge funding from the European Research Council, via
the award of an ERC Consolidator Grant (P.I. R. McLure). VAB, JSD, and RAAB acknowledge the support of the
European Research Council through the award of an Advanced Grant (P.I. J. Dunlop). SP acknowledges the support of the University of Edinburgh via the Principal's Career Development Scholarship. VAB and JSD acknowledge the support of the EC FP7 Space project ASTRODEEP (Ref. No: 312725). RAAB acknowledges the support of the Oxford Centre for Astrophysical Surveys which is funded through generous support from the Hintze Family Charitable Foundation. We thank Shy Genel, Bruno Henriques, Violeta Gonz\'{a}lez-P\'{e}rez, and Romeel Dav\'{e} for providing their simulations for comparison and for useful discussion. This work is based in part on observations made with the NASA/ESA \textit{Hubble Space Telescope}, which is operated by the Association of Universities for Research in Astronomy, Inc, under NASA contract NAS5- 26555. This work is based on data products from observations made with ESO Telescopes at the La Silla Paranal Observatories under ESO programme ID 179.A2005 and on data products produced by TERAPIX and the Cambridge Astronomy survey Unit on behalf of the UltraVISTA consortium. This work is based in part on observations made with the \textit{Spitzer Space Telescope}, which is operated by the Jet Propulsion Laboratory, California Institute of Technology under a NASA contract.
\bibliographystyle{mnras}
\bibliography{refs}

\label{lastpage}
\end{document}